\documentclass[pmlr,twocolumn,10pt]{jmlr} 

\usepackage{booktabs}
\usepackage{siunitx}

\usepackage{times}

\usepackage{soul}
\usepackage{url}
\usepackage{enumitem}

\newcommand{\equal}[1]{{\hypersetup{linkcolor=black}\thanks{#1}}}

\theorembodyfont{\upshape}
\theoremheaderfont{\scshape}
\theorempostheader{:}
\theoremsep{\newline}

\jmlryear{2022}
\jmlrworkshop{Conference on Health, Inference, and Learning (CHIL) 2022} 

\title[PhysioMTL]{PhysioMTL: Personalizing Physiological Patterns using Optimal Transport Multi-Task Regression}

\author{%
\Name{Jiacheng Zhu\textsuperscript{$\ddagger$}}\equal{Work done while at Apple.} 
\Email{jzhu4@andrew.cmu.edu}\\
\Name{Gregory Darnell\textsuperscript{$\dagger$}}
\Email{gdarnell@apple.com}\\
\Name{Agni Kumar\textsuperscript{$\dagger$}} 
\Email{agni@apple.com}\\
\Name{Ding Zhao\textsuperscript{$\ddagger$}} 
\Email{dingzhao@cmu.edu}\\
\Name{Bo Li\textsuperscript{$\mathsection$}} 
\Email{lbo@illinois.edu}\\
\Name{Xuanlong Nguyen\textsuperscript{$\mathparagraph$}} 
\Email{xuanlong@umich.edu}\\
\Name{Shirley You Ren\textsuperscript{$\dagger$}} 
\Email{shirleyr@apple.com}\\
\addr
\textsuperscript{$\dagger$,*}Apple,
\textsuperscript{$\ddagger$}Carnegie Mellon University,
\textsuperscript{$\mathsection$}UIUC, 
\textsuperscript{$\mathparagraph$}University of Michigan
}
  
\begin{document}

\maketitle
\begin{abstract}
Heart rate variability (HRV) is a practical and noninvasive measure of autonomic nervous system activity, which plays an essential role in cardiovascular health.
However, using HRV to assess physiology status is challenging. Even in clinical settings, HRV is sensitive to acute stressors such as physical activity, mental stress, hydration, alcohol, and sleep. Wearable devices provide convenient HRV measurements, but the irregularity of measurements and uncaptured stressors can bias conventional analytical methods.
To better interpret HRV measurements for downstream healthcare applications, we learn a personalized diurnal rhythm as an accurate physiological indicator for each individual.
We develop Physiological Multitask-Learning (PhysioMTL) by harnessing Optimal Transport theory within a Multitask-learning (MTL) framework. The proposed method learns an individual-specific predictive model from heterogeneous observations, and enables estimation of an optimal transport map that yields a push forward operation onto the demographic features for each task.
Our model outperforms competing MTL methodologies on unobserved predictive tasks for synthetic and two real-world datasets.
Specifically, our method provides remarkable prediction results on unseen held-out subjects
given only $20\%$ of the subjects in real-world observational studies.
Furthermore, our model enables a counterfactual engine that generates the effect of acute stressors and chronic conditions on HRV rhythms.

\end{abstract}

\paragraph*{Data and Code Availability}

The study in this paper uses two real-world datasets: (1) the \textit{Apple Heart \& Movement Study (AHMS)} dataset~\citep{AHMS_dataset} and (2) the \textit{Multilevel Monitoring of Activity and Sleep in Health people (MMASH)} dataset~\citep{rossi2020public_mmash}. The AHMS dataset is not publicly available.
The MMASH dataset is publicly available on the PhysioNet repository~\citep{MMASH_PhysioNet}, with an open-source python API available at \href{https://github.com/RossiAlessio/MMASH}{https://github.com/RossiAlessio/MMASH}. We will make code available to reproduce all the experimental results obtained from the MMASH dataset.

\section{Introduction}
\label{sec:intro}

Heart Rate Variability (HRV) \citep{acharya2006heart_hrv_review,electrophysiology1996_hrv} is a reliable measure used in physiological and psychological research, since HRV reflects cardiac autonomic nervous system (ANS) regulation variation (i.e., the changing balance between the sympathetic and parasympathetic nervous systems).
Beyond a purely cardiovascular measure, HRV has been used as a tool to detect acute illness (e.g., early detection of COVID-19~\citep{warrior_watch}), common cold~\citep{grzesiak2021assessment}, and inflammatory response from infection~\cite{hasty2021heart}).
HRV also aids diagnostics~\citep{acharya2006heart_hrv_review}, and is predictive of various cardiovascular-related diseases~\citep{zhang2016association}.
Numerous studies have investigated the connection between HRV and a range of conditions, including cardiovascular \citep{kamath1993power_hrv_cardiac,akselrod1981power_hrv_cardio,berger1986efficient_hrv_cardio}, blood pressure \citep{de1985relationships_bloodpressure,akselrod1985hemodynamic_bloodpressure}, and myocardial infarction \citep{duru2000effect_MyoInfa,katona1975respiratory_MyoInfa,rothschild1988temporary_MyoInfa}.

However, the assessment of cardiovascular functioning using HRV is challenging in practice. HRV is sensitive to many acute stressors, including alcohol, exercise, sickness~\citep{altini2021behind}, and hydration~\citep{christiani2021cardiac}.
To remove confounding factors, most clinical studies set strict rules for HRV measurements~\citep{pop2021assessment_HRV_alcohol} (e.g. in dorsal decubitus, in a quiet and temperature-controlled room, and assessment in the morning or right after awakening). The increase in wearable devices has enabled tracking cardiovascular status more conveniently, which motivates larger-scale scientific studies~\citep{warrior_watch,bowman2021method_michigan}. Yet, measurement missingness can be non-random, making it challenging to apply a unified statistics approach (e.g., average) for HRV analysis.

In this context, machine learning (ML) techniques open the realm to deal with noisy, irregularly sampled physiological data. 
Specifically, multi-task learning (MTL)~\citep{ruder2017overview_mtl,zhangyang2018overview_mtl} is relevant when we consider the objective of modeling each individual’s cardiovascular status as a separate task rather than learned a model pooled across all individuals. 
There is an increasing interest in applying MTL in the medical and healthcare domain. 
For example, an Ontology-driven MTL framework \citep{ghalwash2021phenotypical_mtl_chil} considers different cohorts of phenotypes as various tasks, and improves prediction performance with a representation sharing technique guided by structure information. 
To better incorporate the relationships between covariates of electronic health records (EHR), a multi-task Gaussian Process (GP)~\citep{cheng2020sparse_mtl_gp} framework is proposed for high-quality inference of different disease subgroups and studies. Multi-task LSTM models~\citep{kumar2021estimating_mtl} are shown to be effective in the remote estimation of Respiratory rate (RR). 
A Wasserstein multi-task regression framework (MTW)~\citep{janati2019wasserstein_mtl,janati2020multi} that exploits geometric information is shown effective in inferring EEG signals collected from multiple subjects. 
Also, MTL methods are extremely helpful in the assessment of Parkinson's disease~\citep{shujianyu2020learning_graph_mtl} and aging~\citep{zhou2011multi_jiayu_diesease}.

\begin{figure}[t]
  \centering
  \includegraphics[width=0.95\linewidth]{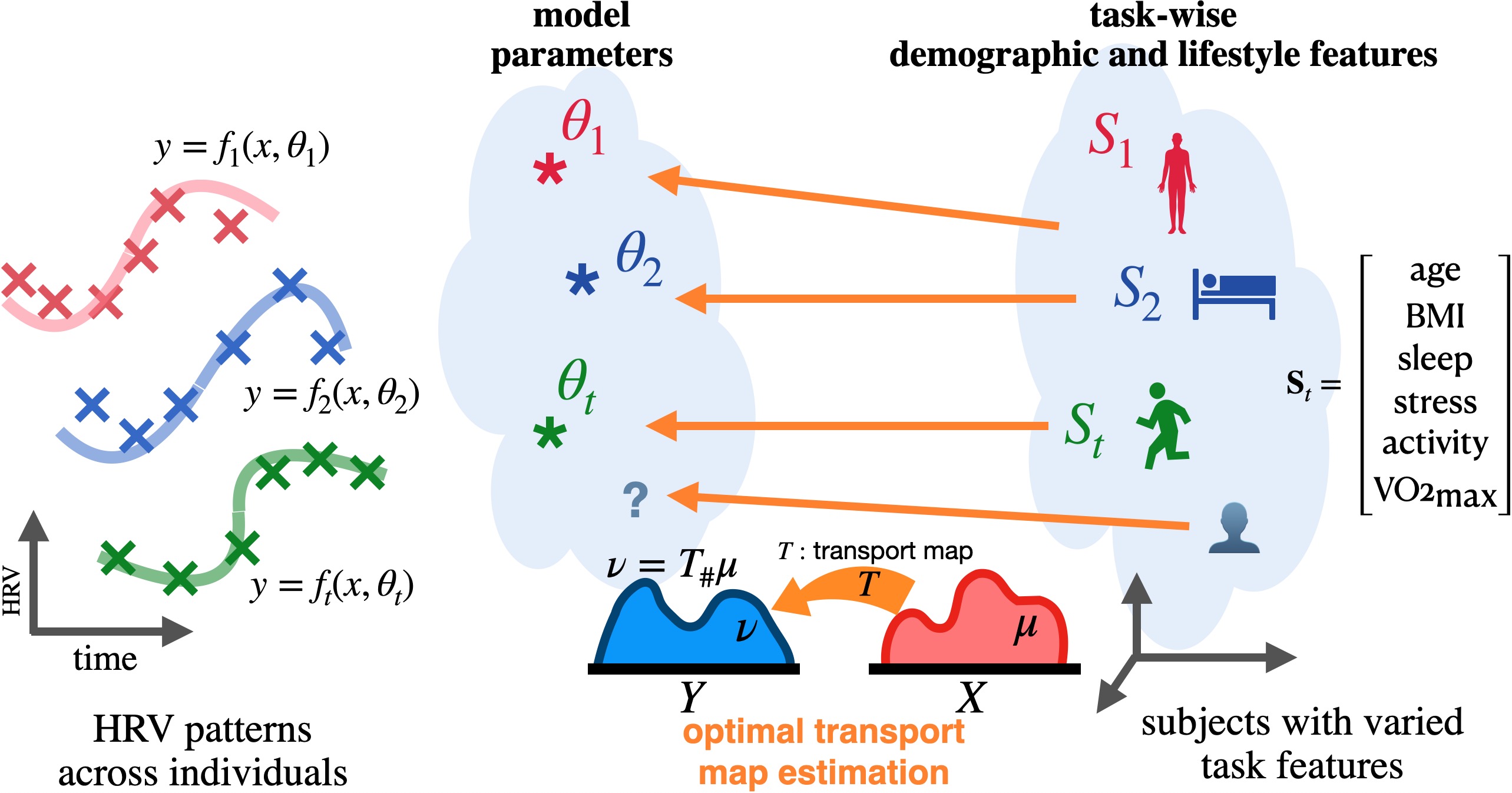}
  \label{fig:first_figure}
  \vspace{-10pt}
  \caption{
  A schematic of our learning method and tasks.
  }
\end{figure}

Furthermore, clinical and healthcare data usually contain significant heterogeneity in demographics, treatments, and devices,
which makes \textit{domain generalization} a critical problem. 
Usually, clinical ML models experience degraded performance in \textit{out-of-distribution} datasets that are not seen during training~\citep{zhang2021empirical_domain_gen_chil2021}.

In this work, \textbf{we design a methodology with the following capabilities}: 
(1) can capture the latent physiological trajectory from sparse and noisy HRV measurements,
(2) is personalized per subject, since baseline cardiovascular control  varies among individuals---the genetic heritability of HRV may be up to 71\%~\citep{nolte2017genetic}---and (3) can generalize from acute and chronic covariates in a training dataset to unseen subjects that may not be sampled from the training distribution. An illustration of our method is given in figure (\ref{fig:first_figure}).
In summary, our contribution is threefold: 
\begin{itemize}
\itemsep0pt      
    \item To obtain a personalized HRV baseline, we extend the physiology-informed circadian rhythm regression problem~\citep{jarczok2019circadian_sinusoi,berger1986efficient_hrv_cardio} to the multi-task learning setting, where we comprehensively assess the relatedness between physiological variation and individual information such as demographics, activity, and acute stress.
    \item We propose to handle domain-generalization using optimal transport (OT) map estimation methods.  We estimate a domain-invariant mapping that is generalizable for unseen out-of-domain task distributions.
    \item We validate our proposed methodology on synthetic and real-world datasets that contain comprehensive physical, physiological, and psychological health records:  the Apple Heart and Movement Study (AHMS) dataset and the Multilevel Monitoring of Activity and Sleep in Heathy people (MMASH) dataset~\citep{rossi2020public_mmash}.
\end{itemize} 
Our proposed method demonstrates state-of-the-art performance on the out-of-sample testing tasks that are not available in the training stage, which indicates our approach will be effective for out-of-distribution domain generalization.
Moreover, we use the estimated transport map to investigate the effects of cardiovascular risk factors. Specifically, we perform a counterfactual analysis by generating the circadian rhythm variationals caused by different risk factors.

\section{Background \& Related Work}
\label{gen_inst}

\textbf{Multi-task learning.}
Recent survey papers~\citep{ruder2017overview_mtl,zhangyang2018overview_mtl} provide a comprehensive view of MTL research.
For multi-task regression problems, one classical way to exploit task relatedness is the sparsity assumption of regression coefficients, which has lead to a family of Lasso-type models~\citep{evgeniou2007multi_mtl_feature,obozinski2006multi_lasso}.
Also, a temporal group Lasso regularizer is proposed to improve the task generalization performance for disease progression prediction~\citep{zhou2011multi_jiayu_diesease}. 
When task-wise features are available, task-similarity can be encoded by kernel functions \citep{bonilla2007kernel,bickel2008multi_kernel_hiv} and serve as a prior. 
To achieve better generalization and adaptation for MTL algorithms, \citep{wang2021bridging} showed MTL is closely related to gradient-based Meta-learning~\citep{finn2017model_maml} from an optimization perspective and could adapt to unseen tasks. 
There have been a few studies that use optimal transport theory and the Wasserstein distance for MTL generalization. 
An unbalanced Wasserstein distance is applied~\citep{janati2019wasserstein_mtl} to compute the distance between task parameters with non-overlapping support. Wasserstein distance also provides benefits in an adversarial multi-task neural network model~\citep{shui2019principled_mtl}. 
In addition, OT has demonstrated an advantage in dealing with covariate shift for MTL tasks~\citep{liu2019augmented_mtl_ot}.

\textbf{ML for HRV circadian rhythms.}
Traditionally, time-domain analysis~\citep{electrophysiology1996_hrv}, frequency-domain analysis~\citep{kim2009effect_frequency,clifford2005quantifying_frequency_spectral}, and nonlinear analysis~\citep{west2012fractal_physiology_nonlinear,akay2000nonlinear_biomedical} are the primary analysis methods for obtaining features from HRV signals.
Not until recently have modern ML approaches been applied to HRV. One pioneer study used a multilayer perceptron~\citep{liu2014utility_hrv_ml_first} to identify lifesaving interventions (LSIs) in trauma patients with HRV signals. Deep neural networks (NNs) are also effective for Myocardial Infarction (MI) detection using HRV data~\citep{shahnawaz2021effective_hrv_ml}. HRV studies also involve other ML methods including K-nearest neighbors (KNN)~\citep{narin2018early_hrv_ml_knn}, support vector machine (SVM)~\citep{shi2019renyi_hrv_ml_svm} and convolutional nerual networks (CNN)~\citep{taye2020application_hrv_ml_cnn}.
Recently, deep sequential models~\citep{ni2019modeling_heart_rate_fitness} and transfer learning approaches~\citep{spathis2021self_physio_transfer_CHIL,spathistowards_physio_transfer} have been proposed to assess other physiological features, such as Heart Rate (HR), from wearable devices.
Nevertheless, to the best of our knowledge, there are no studies that apply MTL to personalize HRV patterns.

To reveal the underlying physiological status for each task, we incorporate prior physiological knowledge. Circadian rhythms are ubiquitous in almost any biosignal~\citep{astiz2019mechanisms_circadian_mammal}, and ANS activity fluctuates in a diurnal variation pattern~\citep{jarczok2019circadian_sinusoi} with an approximate solar day frequency, with the peak values occurring during nighttime~\citep{huikuri1994circadian_hrv_nightpeak,li2011circadian_hrv_nightpeak,jarczok2013heart_hrv_nightpeak}. Therefore, we can estimate a sinusoidal pattern~\citep{jarczok2019circadian_sinusoi} from raw HRV measurements and use the variation of pattern parameters as an estimate of the true physiological status regardless of measurement timing. 
An HRV rhythm function is particularly useful for wearables and ambulatory monitoring since they capture real-time \textit{continuous} measurements.

\section{Learning Methods}
\label{methods}

\subsection{Multi-Task regression}
Let us consider a dataset with $T$ tasks $\{X_t, Y_t, S_t\} = \{ (x_{ti}, y_{ti}, s_{t}) : i \in \{ 1,...,N_t\}\}$, where each task $t \in \{ 1,.., T\}$. $X_t \in \mathbb{R}^{d_x \times N_t}$ contains $N_t$ samples and $d_x$ features, while $Y_t \in \mathbb{R}^{1 \times N_t}$ is the label for task $t$ and $s_t \in \mathbb{R}^{d_s \times 1}$ is the task-wise feature vector. We start from multi-task linear regression with $\mathbf{W} = [W^T_1, ...,W^T_T] \in \mathbb{R}^{T \times d_w}$, the MTL optimization is formulated as:
\begin{equation}
\label{eq:mtl_first}
    \min_{\mathbf{W}} \frac{1}{2} \sum_{t=1}^T \| W^T_t X_t  - y_t\|^2_2 + \Omega(\textbf{W})
\end{equation}
where the second term is a regularizer that exploits the task similarities and/or desired parameter sparsity. For example, in MTLasso~\citep{obozinski2006multi_lasso}, the regularization is 
\begin{equation}
    \Omega(W) = \lambda_1 \| \mathbf{W}\|_1 + \lambda_2 \| \mathbf{W} \|_F
\end{equation}
where $\lambda_1$ specifies the task sparsity and $\lambda_2$ controls the model complexity. The relationship between tasks can also be formulated as the similarity matrix or the adjacency matrix of a graph~\citep{he2019efficient_ccmtl,alesiani2020towards_graph_mtl,shujianyu2020learning_graph_mtl} such as:
\begin{eqnarray}
\Omega(W) = \sum_{i=1}^T \sum_{j \in \mathcal{N}_i} A_{ij} \| W_i -  W_j\|^2_2
\end{eqnarray}
where $A$ encodes the relatedness between tasks and $\mathcal{N}_i$ is the set of nearest neighbors of $i$.

One of the objectives of our study is that we not only want to obtain predictive functions for each training task but also perform predictions for those tasks that are not available during training. 
Specifically, with a dataset  $\{X_t, Y_t, S_t\}_{t=1}^T$ collected from $T$ training tasks $\{ \mathcal{T}_t \}_{t=1}^T$, we aim at the prediction for $T'$ new tasks $\{ \mathcal{T}_t^{new} \}_{t=1}^{T'}$ where only task-wise features $\{S^{new}_t\}_{t=1}^{T'}$ are available. 
In our case, to address this problem, we assume each task $\mathcal{T}_t$ is associated with a set of predictive function model parameters $W^\top_t$ and a task-wise feature $S_t$, which are actually drawn from distributions $S_t \sim \mu_S$ and $W^\top_t \sim \nu_W$, respectively. 
A promising idea is to exploit the similarity between these two probability measures $\nu_S$ to $\mu_W$, and to obtain a predictive transformation that transports every unit of mass from one probability measure to another. To deal with probability distributions, optimal transport (OT) theory~\citep{villani2009optimal_old_new} is a canonical choice. 

\subsection{Optimal Transport and Map Estimation}
Let $\mathcal{X} \subseteq \mathbb{R}^{d_x}$ and $\mathcal{Y} \subseteq \mathbb{R}^{d_y}$ be two complete and separable metric spaces and $\mathcal{M}(\mathcal{Z})$ denote the space of probability measures over spaces $\mathcal{Z}$.
Given two probability measures $\mu \in \mathcal{M}(\mathcal{X})$, $\nu \in \mathcal{M}(\mathcal{Y})$ and a cost function $c: \mathcal{X} \times \mathcal{Y} \mapsto \mathbb{R}^+$, the Monge problem consists in finding a Borel map, $T: \mathcal{X} \mapsto \mathcal{Y}$ between $\mu$ and $\nu$. 
Monge's formulation can be improved by the Kantorovich formulation which seeks a joint measure $\pi \in \Pi$ by minimizing 
 
\begin{eqnarray}
\label{eq:Kantorovich}
\pi^* = \arg \min_{\pi \in \Pi} \int_{ \mathcal{X} \times \mathcal{Y}} c(x, y) d \pi(x, y).
\end{eqnarray}
Here, $\Pi =\{\pi: \gamma_\#^{\mathcal{X}} \pi = \mu, \gamma_\#^{\mathcal{Y}} \pi = \nu\}$ is the probabilistic couplings in the space of joint distributions with marginals $\mu$ and $\nu$.
Note that the optimal coupling $\pi^*$ always exists and the conditional probability distributions $\pi_{y|x}$ gives stochastic maps from $\mathcal{X}$ to $\mathcal{Y}$ following \textit{barycentric mapping}. 
However, the main limitation of the barycentric mapping is that it can not be applied to project \textit{out-of-sample} examples which are \textit{not observed} during the learning process of $\pi^*$. 
To that end, it is helpful to estimate the transport map with a transformation function~\citep{perrot2016mapping,seguy2017large_LSOT,makkuva2020optimal_ICNN,manole2021plugin_map_rate,zhu2021functional_fot}.

\subsection{PhysioMTL with OT map estimation}
We assume the task-wise feature vectors $S_t$ and their underlying predictive functions (parameters) $W^\top_t$ are governed by probability measures $\mu_S \in \mathcal{M(S)}$ and $\nu_W \in \mathcal{M(W)}$, where $\mathcal{S} \subseteq \mathbb{R}^{d_s}$ and $\mathcal{W} \subseteq \mathbb{R}^{d_t}$ are two metric spaces. We want an optimal transport map $T$ between $\mu_S$ and $\nu_W$ such that $T_\# \mu_S = \nu_W$, or equivalently, $T(S) = W^\top \sim \nu_W$ for whatever $S \sim \mu_S$.
Hence, we propose to estimate this transport map with some transformation function $F: \mathbb{R}^{d_s} \mapsto \mathbb{R}^{d_w}$
with:
\begin{equation}
\label{eq:map_est_objective}
    \hat{F} = \arg \min_{\pi \in \Pi, F \in \mathcal{F}} \int_{\mathcal{S} \times \mathcal{W}} c(F(s), w) d \pi (F(s), w).
\end{equation}
Note that joint estimation of the coupling and map, as equation~(\ref{eq:map_est_objective}) does, is shown to be effective in a few OT studies~\citep{alvarez2019towards,rangarajan1997softassign}, but this is the first time optimal transport map estimation is used in a multi-task regression.

In the practical setting, we could only observe samples from $\mu_S$ and $\nu_W$.
So we associate empirical measures $\hat{\mu}_s =\sum_{t=1}^{T} p_t^s \delta_{s_t}$ and $\hat{\nu}_w =\sum_{t=1}^{T} p_t^w \delta_{W_t}$ with  task-wise features $\{\mathbf{s}_t\}_{t=1}^T$ and the parameters to be estimated $\{W_t\}_{t=1}^T$. Then we propose a new regularization for multi-task learning as
\begin{equation}
\label{eq:ot_mtl_reg}
    \Omega(W) = \sum_{i,j=1}^T \pi_{i,j} c(F(\mathbf{s}_i), W_j) 
\end{equation}
Where $c: \mathcal{W} \times \mathcal{W} \mapsto \mathbb{R}^+$ is the cost function and $\pi \in \mathbb{R}^{T \times T}$ is the probabilistic coupling.

This regularization term has several interpretations: 
(1) The OT formulation’s flexibility allows us to make minimum assumptions on the distribution of task-wise features and predictive functions and only rely on empirical distributions, which is critical for modeling complex physiological indicators such as HRV.
(2) Our map estimation formulation with a continuous transformation function allows similar task-wise features to predict similar model parameters, as is the case in physiology studies~\citep{pop2021assessment_HRV_alcohol,jarczok2019circadian_sinusoi}.
(3) Inspired by \cite{janati2019wasserstein_mtl}, we use the Wasserstein distance to measure the similarity of model parameters. In addition, we can efficiently use prior knowledge on the geometry of regressors by exploiting the \textit{ground metric} in the OT framework. 
(4) The probabilistic coupling $\pi$ has a clear interpretation since it represents the task relatedness and has a similar effect as a similarity graph~\citep{alesiani2020towards_graph_mtl,shujianyu2020learning_graph_mtl,he2019efficient_ccmtl}.

\subsubsection{Parameterizing the transport map}
Above we presented our proposed regularization with a general set of transformation functions. Now we propose several choices of a convex set $\mathcal{F}$. We first consider a linear transformation, where $\mathcal{F}$ is defined by a matrix $\mathbf{F} \in \mathbb{R}^{d_s \times d_{\theta}}$ representing all the affine transformations 
\begin{equation}
    \mathcal{F} = \{F: \mathbf{F} \in \mathbb{R}^{d_{\theta} \times d_s}, \forall \mathbf{s}_t \in \Omega_S, F(\mathbf{s}_t) = \mathbf{F} {} \mathbf{s}_t  \}
\end{equation}
Usually a linear transformation is not sufficient to approximate the transport map, particularly for modeling more complex systems like the human ANS. Therefore, we consider \textit{non-linear} transformations. Let $\phi$ be a nonlinear function associated to a kernel function $k: \Omega_s \times \Omega_s \mapsto \mathcal{R}$ s.t. $k(s_i, s_j) = < \phi(s_i), \phi(s_j)>$, then for a given set of samples $\mathbf{S}$ we can define $\mathcal{F}$ as
\begin{equation}
    \mathcal{F} = \{F: \mathbf{F} \in \mathbb{R}^{d_{\theta} \times T}, \forall \mathbf{s}_t \in \Omega_S, F(\mathbf{s}_t) = \mathbf{F} {} \mathbf{k}_{\mathbf{s}_t}(\mathbf{s}_t) \}
\end{equation}
here $\mathbf{k}_{\mathbf{s}_t}( \cdot )$ denotes the vector $(k(\mathbf{s_1}, \cdot), ..., k(\mathbf{s_T}, \cdot))$. 

\begin{figure*}[htbp]
\floatconts
  {fig:subfigex}
{\caption{Qualitative illustration of PhysioMTL on simulated dataset (best viewed in color). Here each task is marked by color according its task-specified feature visualized in \textbf{(b)}. 
In \textbf{(a)}, We can see that the curves are varying smoothly with respect to this feature. In \textbf{(b)}, we visualize the distribution of both training and testing tasks as $u(s_t)$ and $v(s_e)$ through \textit{kernel density estimation (KDE)}, and we can see the testing tasks have a small overlap with the training tasks. In \textbf{(c)}, we show the prediction result of several baselines and our method. While other methods struggles to provide accurate prediction, our method (PhysioMTL) effectively generalizes to the out-of-sample unseen testing tasks. }}
  {%
    \subfigure[Training tasks]{\label{fig:toy_train}%
      \includegraphics[width=0.416\linewidth]{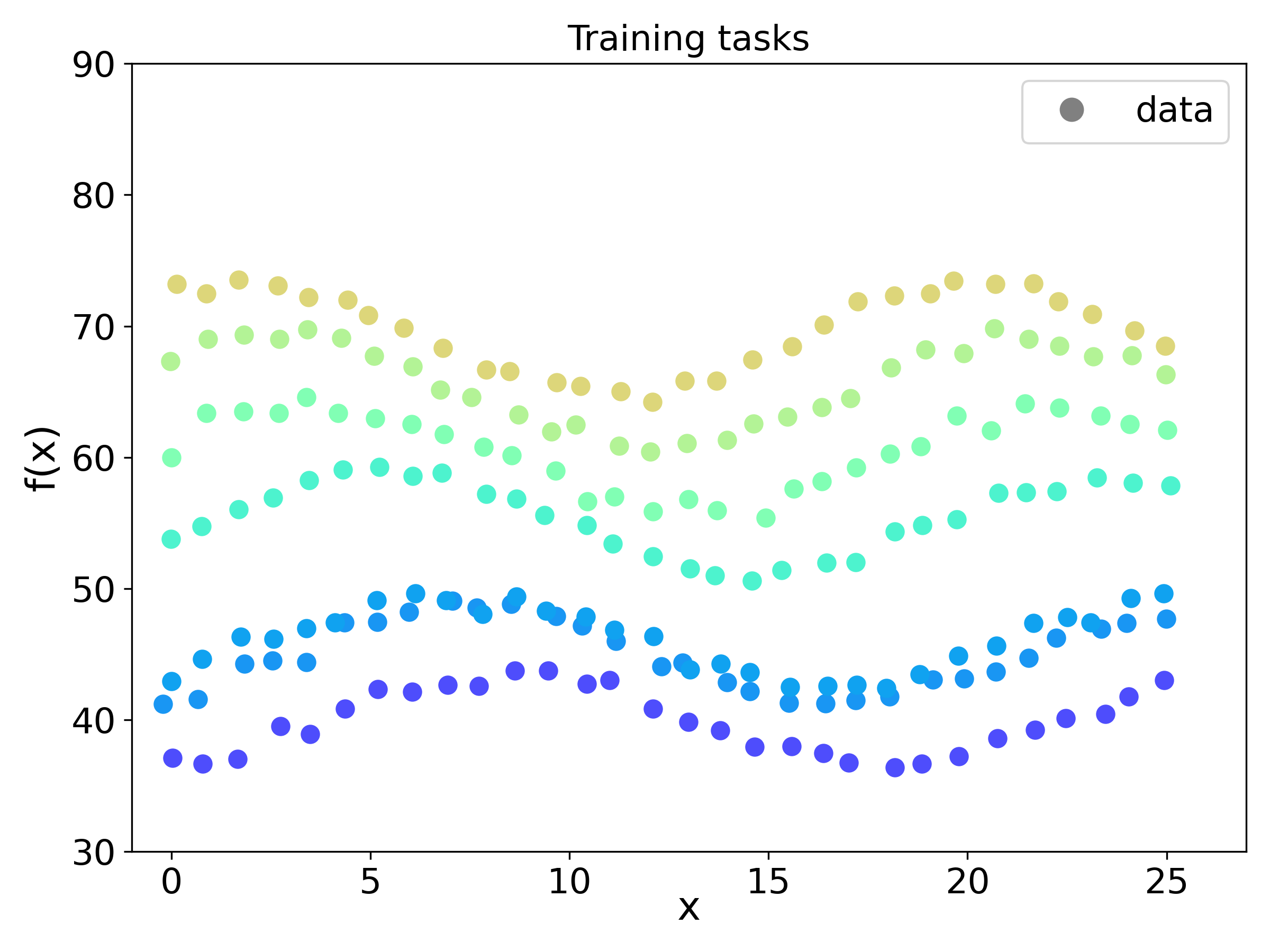}}%
    \centering
    \subfigure[Relatedness]{\label{fig:toy_relatedness}%
      \includegraphics[width=0.161\linewidth]{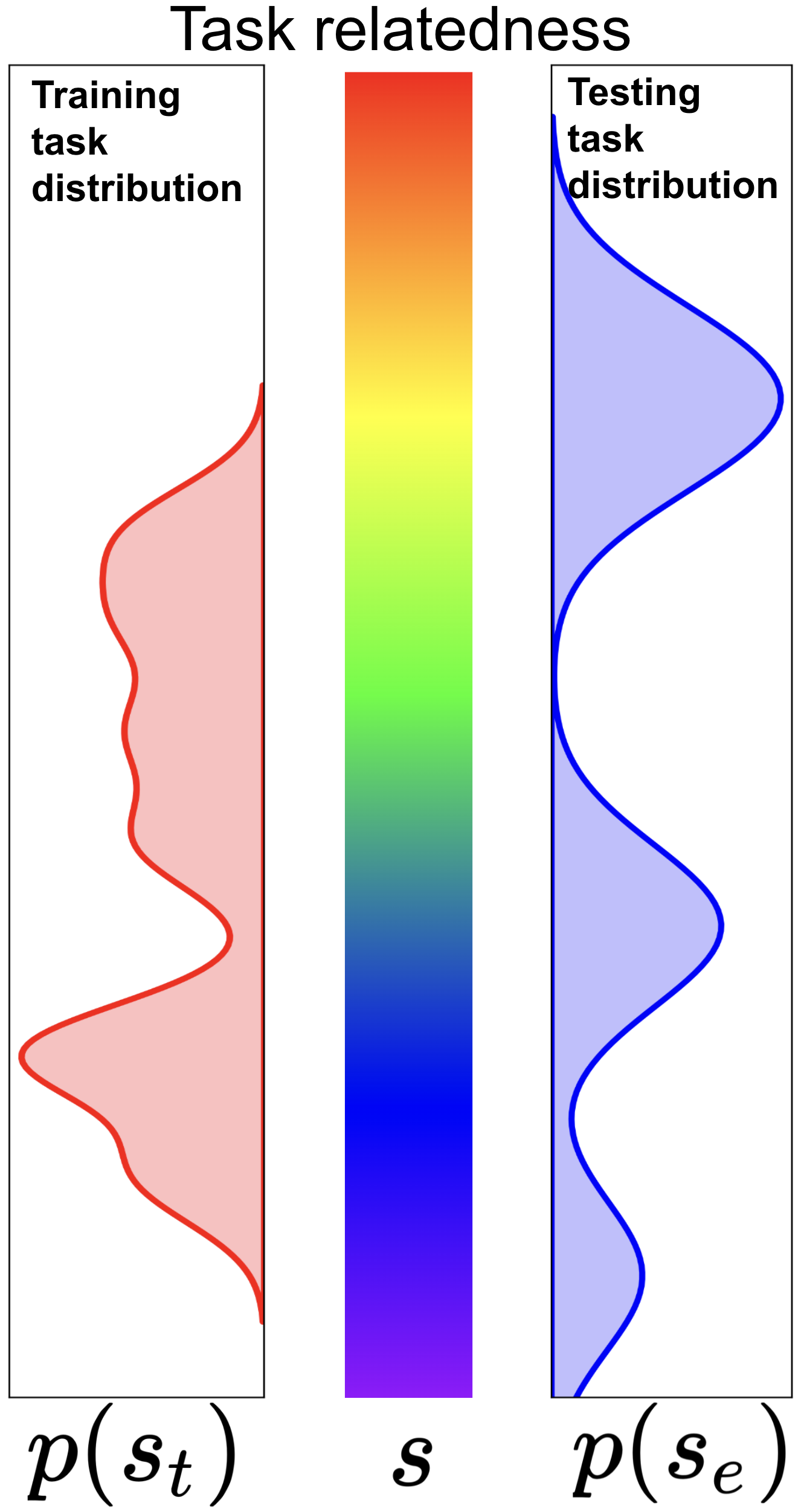}}
    \subfigure[Test on unseen tasks]{\label{fig:toy_test}%
      \includegraphics[width=0.416\linewidth]{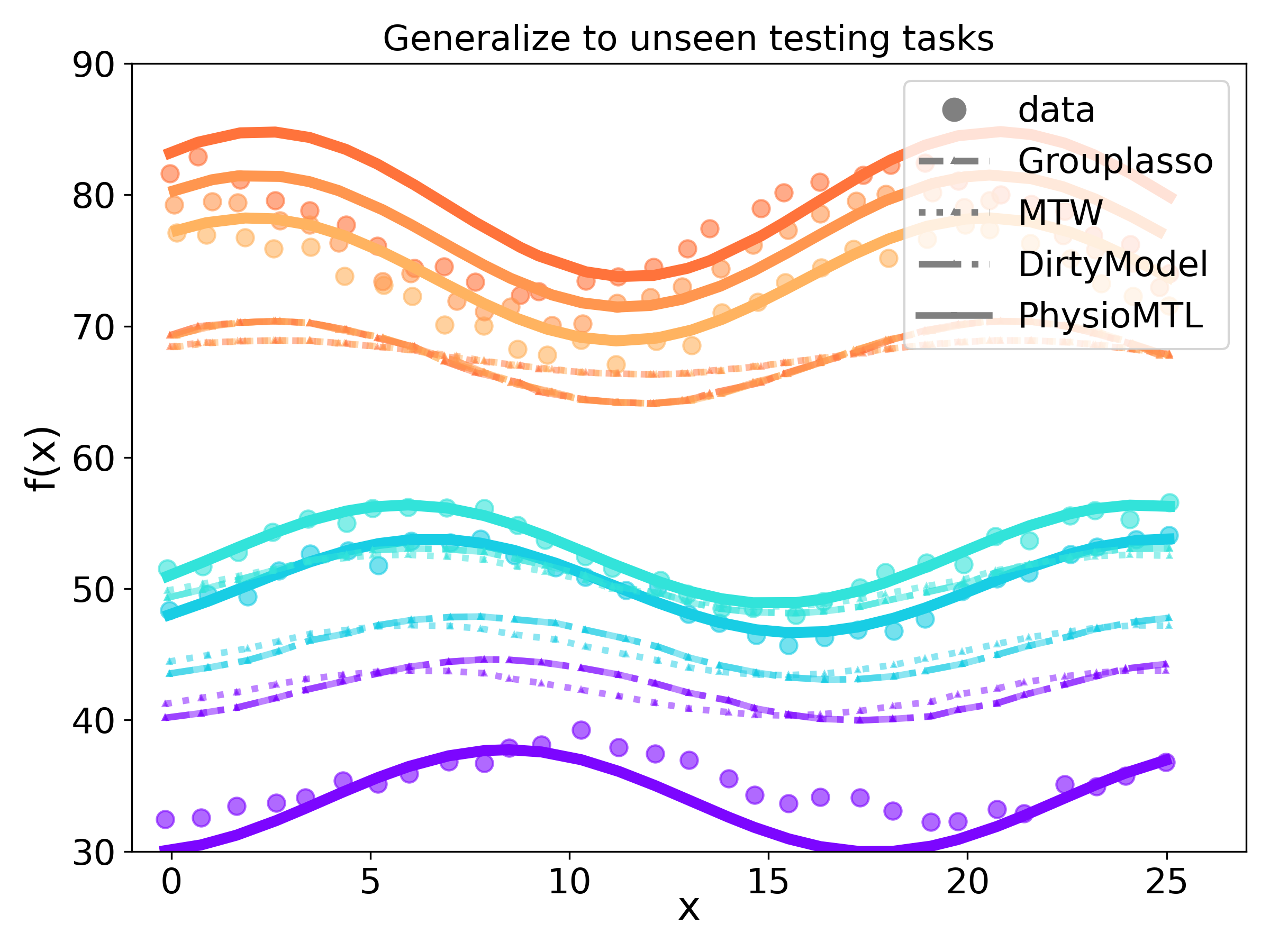}}
  \vspace{-30pt} 
  }
\end{figure*}

\subsection{Modeling HRV circadian rhythms}

In practice, HRV data collected with mobile devices are unevenly distributed across time, thus challenging the application of most sequential machine learning models that assume evenly spaced data.
Physiological studies ~\citep{jarczok2019circadian_sinusoi,berger1986efficient_hrv_cardio} have suggested to model the nonlinear circadian rhythm with a sinusoidal pattern (e.g., a cosinor model). 
For an individual $t$, the daily HRV circadian rhythm over a 24 hour period can be written as the following nonlinear function:
\begin{equation}
\begin{split}
    \label{eq:sinu}
    hrv(\tau) = M_{t} + A_{t} \sin ( \frac{2 \pi \tau}{P} + \phi_{t}) + \epsilon_{t}(\tau) \smallskip  
\end{split}
\end{equation}
where $M_{t}$ is the Midline Statistic of Rhythm (MESOR), $A_{t}$ is the amplitude, $\phi_{t}$ is the phase shift and $\epsilon_{t}$ is the noise. 
Assuming the period $P$ is fixed, the model in Eq.(\ref{eq:sinu}) can be transformed into a linear model by expanding the sine term into $x_{1}=\sin(\frac{2 \pi}{P} \tau)$, $ x_{2}=\cos(\frac{2 \pi}{P} \tau)$, yielding
\begin{equation}
\begin{split}
    hrv(\tau) = w_{0t} + w_{1t} x_{1} + w_{2t} x_{2} + \epsilon(\tau), \\
\end{split}
\end{equation}
where $w_{0t} = M_{t}$, $w_{1t} = A_{t}\cos (\phi_{t})$, $w_{2t} = A_{t} \sin (\phi_{t})$.
Given raw HRV measurements $\{h_{t,i}, \tau_i \}_{i=1}^{n_t}$ of one subject $t$, we can 
learn parameters of the circadian rhythm model as a linear regression by rewriting $\mathbf{X}_t=[\mathbf{x}_{t,1}^\top, ..., \mathbf{x}_{t,n_t}^\top] \in \mathbb{R}^{n_t \times 3}$ where $\mathbf{x}_{t,i} = [\sin(\frac{2 \pi}{P} \tau_i), \cos(\frac{2 \pi}{P} \tau_i), 1]$ and $\mathbf{Y}_t = [h_{t,i},..,h_{t,n_t}]^\top \in \mathbb{R}^{n_t \times 1}$.
The benefit of this physiology-informed sinusoidal process is threefold: (1) The estimated sinusoidal parameters can serve as a personalized determination of circadian variation, (2) with a continuous function of time, we can assess underlying ANS status with unevenly sampled HRV measurements, and (3) the linear model naturally fits into a multi-task regression enabling estimation of the relatedness among a population of subjects.

\begin{algorithm2e}[t]
\caption{Joint Learning of $\mathbf{W}$ and $F$}
\label{alg_PhysioMTL}
\KwIn{Data $\{X_t, Y_t, S_t\}$, cost function $C_s(s_i, s_j)$}
\KwOut{$\{ \mathbf{W} = \{W_1, W_2,...,W_t\}, \mathbf{F}\}$}
\For{$t=1$ \KwTo $T$}{
Solve $W_t$ by Linear Regression on $\{X_t, Y_t \}$\;
}
Obtain the cost matrix $\Tilde{\mathbf{C}}_{i,j} = C_s(s_i, s_j)$\;
Solve the OT plan with equation (\ref{eq:sinkhorn})\;
\While{not converge}{
Update $F$ using gradient descent\;
Update $W$ using gradient descent\;}
\end{algorithm2e}

\subsection{Optimization algorithm and computation}
The problem we address is to simultaneously learn the MTL parameters $\mathbf{W}$ and estimate $\pi$ and $\mathbf{F}$ for the associated OT formulation. The objective is given by combining equation~(\ref{eq:mtl_first}) and equation~(\ref{eq:ot_mtl_reg}):
\begin{eqnarray}
\label{eq:obj_mtl_ot}
    \min_{\mathbf{W}, \mathbf{F}, \pi} \frac{1}{2} \sum_{t=1}^T \| W^T_t X_t  - y_t\|^2_2 \nonumber \\
    + \alpha \sum_{i,j=1}^T \pi_{i,j} c(F(\mathbf{s}_i), W_j) 
\end{eqnarray}
The formulation is jointly convex in each variable $\mathbf{W}$, $\mathbf{F}$, and $\pi$ when the remaining are fixed. Therefore, we solve the optimization problem with an alternating minimization framework.

\subsubsection{Entropic OT \& Prior on task similarity}

When fixing $\mathbf{W}$, the second term of equation (\ref{eq:obj_mtl_ot}) is related to a category of OT formulation that can be solved by alternatively minimizing~\citep{alvarez2019towards} over coupling $\pi$ and transformation $\mathbf{F}$. 
Specifically, when solving for $\pi$ with $\mathbf{W}$ and $\mathbf{F}$ fixed, the objective of equation (\ref{eq:obj_mtl_ot}) becomes an $O(n^3 \log n)$ linear program.
Here, adding an entropy regularization will allow us to use the celebrated Sinkhorn algorithm Ag.(\ref{alg:sinkhorn})~\citep{cuturi2013sinkhorn} that just requires $O(n^2)$ operations~\citep{genevay2019_ot_sample_complex}, as: 
\begin{eqnarray}
\label{eq:sinkhorn}
\pi^* = \arg \min_{\pi \in \Pi} \sum_{i,j=1}^T \pi_{i,j} {\mathbf{C}}_{i,j} + \gamma H(\pi),
\end{eqnarray}
where $\mathbf{C}$ is a cost matrix,
$H(\pi) = \sum_{i,j=1}^T \pi_{i,j} \log \pi_{i,j}$ is the negative entropy, and $\gamma$ is a coefficient such that $\gamma \xrightarrow{} 0$ recovers the solution to equation (\ref{eq:Kantorovich}). More details on the Sinkhorn algorithms are in the appendix.

We propose to further ease the computational burden by directly incorporating prior knowledge on the similarity of task-wise features. We pre-compute an optimal transport coupling $\pi^*$ with a cost matrix $\Tilde{\mathbf{C}}_{i,j} = C_s(s_i, s_j)$ whose cost function $c_s: \mathcal{S} \times \mathcal{S} \mapsto \mathbb{R}^+$ is defined as 
\begin{equation}
    c_{s,ij}= \langle \mathbf{m}, | \mathbf{s}_{i}- \mathbf{s}_{j} | \rangle ,
\end{equation}
where $\mathbf{m} \in \mathbb{R}^{d_s \times 1}$ is the similarity coefficient.  
Afterwards, the Sinkhorn algorithm approximates the optimal solution of equation (\ref{eq:Kantorovich}), and we set $\pi = \pi^*$ during the learning procedure. 
This allows us to leverage prior domain knowledge when specifying task similarity (e.g., in our physiological example, we can weight chronic factors higher than acute stressors in our cost function). 
Note that $\mathbf{m}$ is also a potentially learnable parameter~\citep{carlier2020sista_ot_cost,stuart2020inverse_ot_cost}, however we leave the learning of $\mathbf{m}$ to future work.

\subsubsection{Optimization and computation}
With the task-similarity probabilistic coupling $\pi^*$ obtained, our final objective is given by: 
\begin{eqnarray}
\label{eq:obj_fin}
    \min_{\mathbf{W}, \mathbf{F}} \frac{1}{2}  \sum_{t=1}^T  \| W^T_t X_t  - y_t\|^2_2 \nonumber \\ +  \alpha \sum_{i,j=1}^T \pi^*_{i,j} \| F(\mathbf{s}_i)- W_j \|_2^2.
\end{eqnarray} 
Here we use $l_2$ norm as the ground metric for model parameters. We thus solve the objective with Algorithm (\ref{alg_PhysioMTL}), which alternates between minimization of $\mathbf{W}$ and minimization of transport map $\mathbf{F}$. 
Notice that, when instantiating $\mathbf{F}$ to be in the class of linear mappings, minimization with respect to $\mathbf{F}$ has a closed-form solution as it corresponds to an Orthogonal Procrustes problem~\citep{gower2004procrustes}. However, here we are primarily interested in the case where $\mathbf{F}$ is a nonlinear kernel map. Therefore,  
for the chosen parameterization of $\mathbf{W}$, we use gradient descent to solve for the $\mathbf{F}$.
While one could employ an arbitrary kernel, for simplicity we consider the commonly-used Radial Basis Function (RBF) kernel $k(x, x') = \exp \left(- \frac{\|x - x'\|^2}{2 \sigma^2} \right)$ to conduct the kernel trick for a non-linear map. The effectiveness of the non-linear map depends on a proper length scale $\sigma$, and we recommend using cross-validation to select the optimal $\sigma$.
For faster convergence in practice, we initialize $\mathbf{W}$ with prediction models learned independently from each task.
More details on the computation, including the derivation of the gradients, can be found in appendix (\ref{apd:computation}).

\begin{figure}[h!]
  \centering
  \includegraphics[width=0.90\linewidth]{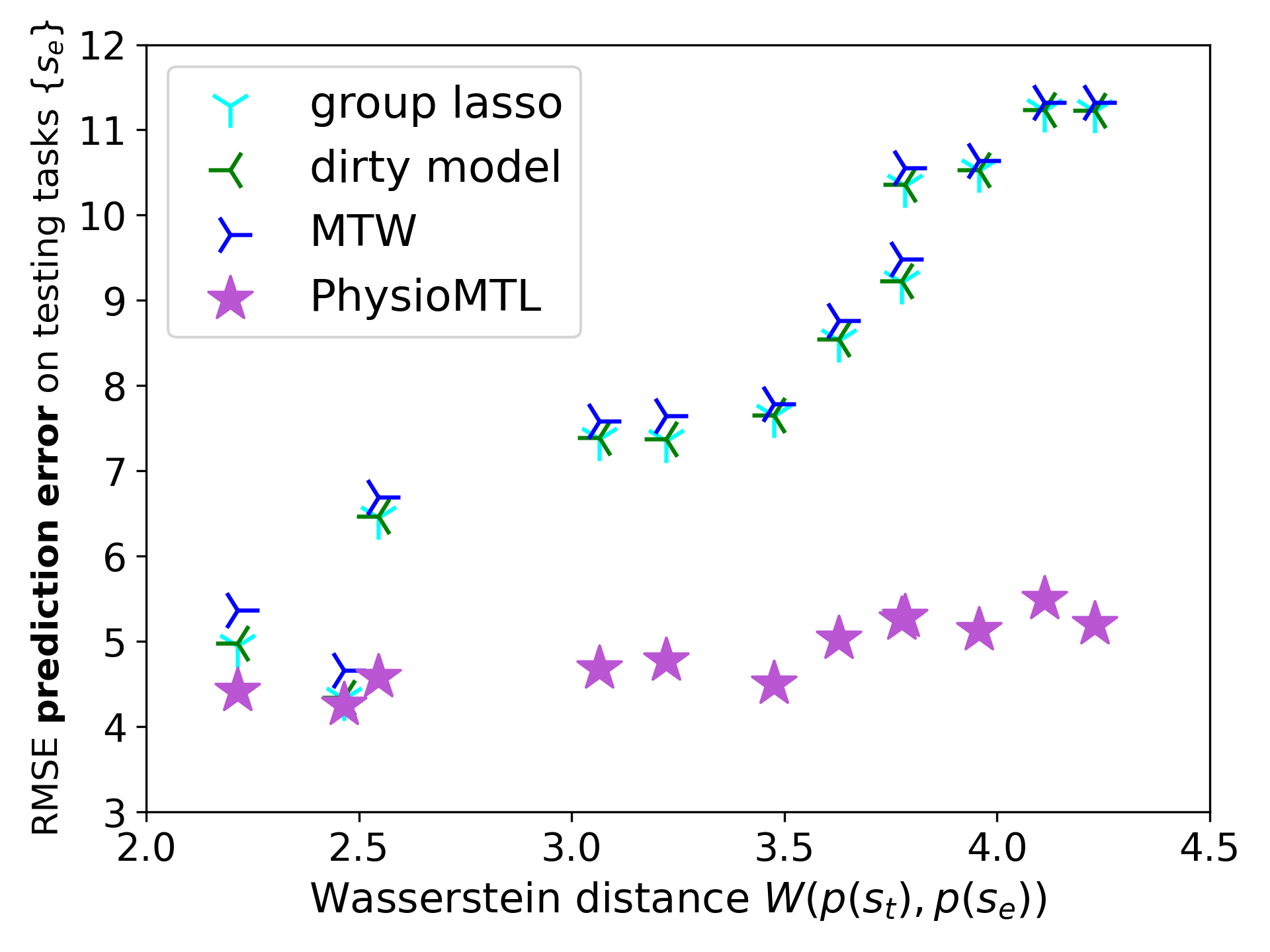}
  \vspace{-12pt}
  \caption{ The prediction error (RMSE) versus the Wasserstein distance between the training and testing tasks $W(\mu(s_t), \nu(s_e))$ with Group Lasso, Dirty models, multi-task Wasserstein (MTW) and our proposed PhysioMTL.}
  \label{fig:toy_example_quant}
\end{figure}

\begin{figure}[h!]
  \centering
  \includegraphics[width=0.99\linewidth]{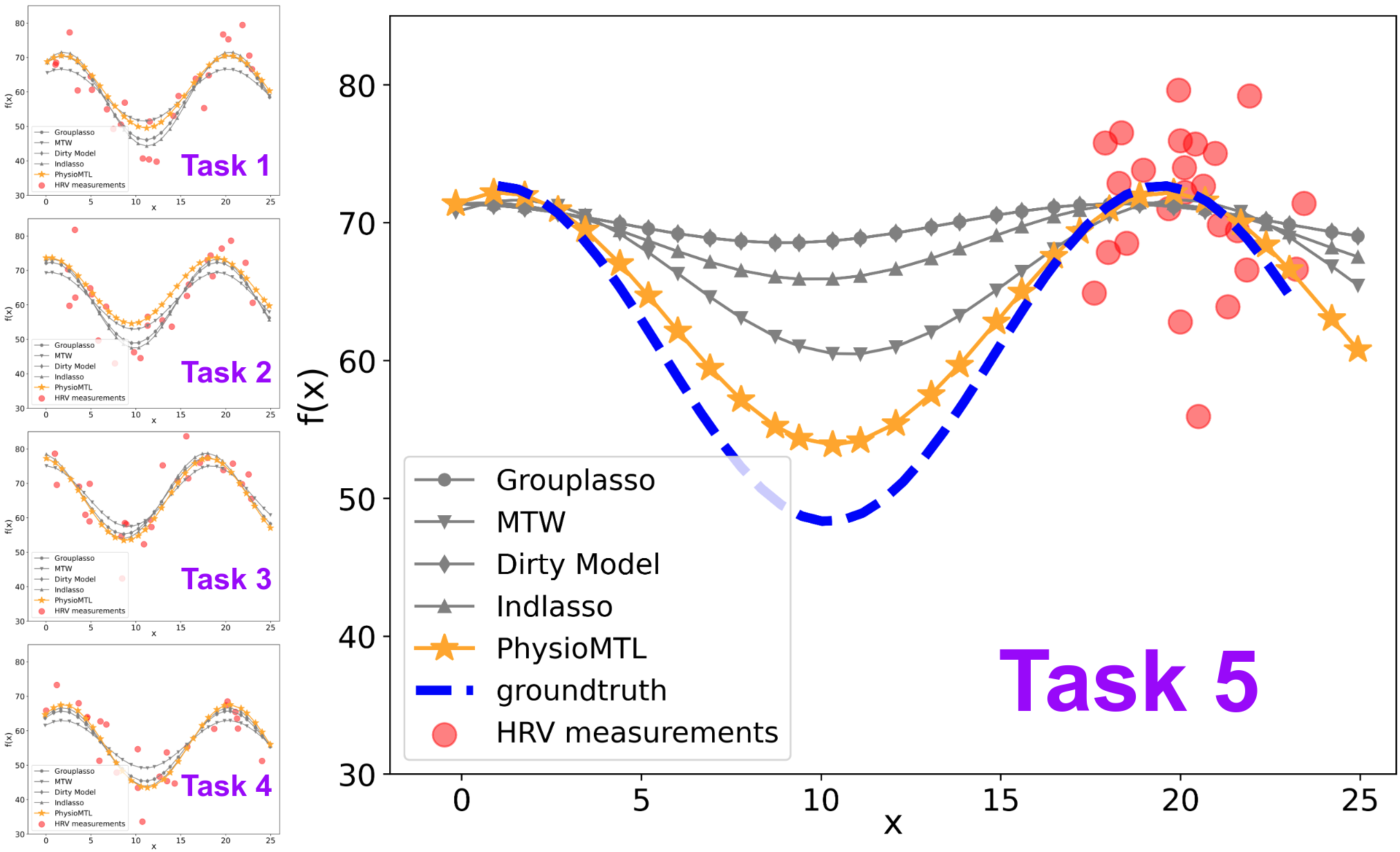}
  \vspace{-10pt}
  \caption{Example of generalization beyond an observed domain when measurements are limited to a particular time period. Tasks one through five are training data for each model; however, task five is missing data from a significant portion of the domain. PhysioMTL extrapolates most accurately to the true latent pattern.
  }
  \label{fig:toy_example_uneven}
\end{figure}

\section{Experiments}
\label{dataset_and_experiments}

In this section, we first verify our proposed PhysioMTL on a synthetic dataset and then apply our method to two real-world datasets. In the first set of experiments, we illustrate how our proposed method adapts to unobserved tasks by comparing the distribution divergence between the training and testing tasks. We also quantitatively verify the advantages of our method over a set of baseline models. Second, we conduct a suite of experiments on two real-world physiology study datasets: the AHMS and MMASH datasets.

In addition to the quantitative evaluations that are standard for machine learning-based methodologies, we show that optimal transport captures meaningful underlying physiology mechanisms and discovers the effects of multiple cardiovascular risk factors from the data. 
There have been studies focusing on the \textit{effects of various cardiovascular risk factors}, including age \& gender~\citep{abhishekh2013influence_hrv_age_gender,voss2015short_hrv_age_gender}, BMI~\citep{koenig2014body_bmi_hrv}, smoking~\citep{murgia2019effects_smoking_hrv}, and anxiety~\citep{chalmers2014anxiety_hrv}. However, most prior work focuses on one or two factors. Comprehensively analyzing multiple factors remains the goal of the current work.
There is also a related and growing interest in applying counterfactual-based causal inference~\citep{prosperi2020causal_health} methods to observational healthcare data. 
Towards this goal, a few pioneer works~\citep{tu2021optimal_causal_ot,li2021causal_ot_treatment} have explored the usage of OT in causal inference.
In our study, we perform a conceptual \textit{counterfactual analysis} by investigating the effect of each factor while holding the remaining fixed. 
To the best of our knowledge, this is the first study that simultaneously reveals the HRV rhythm variation caused by this set of cardiovascular factors (age, BMI, sleep, activity, VO2\textsubscript{max}, and stress level).

\begin{table*}[h!]
\centering
\resizebox{\textwidth}{!}{
\begin{tabular}{lrrrrrr}
\toprule
Method  & hyperparameter & $80\%$ & $60\%$ & $40\%$ & $20\%$ &  \\
\midrule
Global Average       &  n/a  & 38.353 $\pm$ 4.925  & 35.998 $\pm$ 3.793  &  36.941 $\pm$ 1.877 & 36.918 $\pm$ 0.717  \\

Single Lasso                    & $\alpha=0.9$  & 38.306 $\pm$ 5.972  & 35.693 $\pm$ 4.852  & 37.668 $\pm$ 3.110 & 36.187 $\pm$ 1.363   \\

Multi-task Group lasso                    & $\alpha=0.9$  & 38.590 $\pm$ 5.978 & 35.948 $\pm$ 4.815  & 37.928 $\pm$ 3.093 & 36.457 $\pm$ 1.376  \\

Multi-level lasso                    & $\alpha=0.9$  & 38.644 $\pm$ 5.999  & 35.996 $\pm$ 4.810  & 37.999 $\pm$ 3.099 & 36.509 $\pm$ 1.393  \\

Dirty model                    & $\alpha=0.9$  & 38.582 $\pm$ 5.970  & 35.940 $\pm$ 4.818  & 37.921 $\pm$ 3.098 & 36.457 $\pm$ 1.377  \\

MTW                    & $\alpha=0.9$  & 37.896 $\pm$ 5.743  & 35.354 $\pm$ 4.952  & 37.304 $\pm$ 3.084 & 36.170 $\pm$ 1.303  \\
                    
ReMTW                    & $\alpha=0.9$  & 37.941 $\pm$ 5.754  & 35.392 $\pm$ 4.941 & 37.349 $\pm$ 3.082 & 36.187 $\pm$ 1.307   \\
                    
\textbf{PhysioMTL (linear)}       & $\alpha=0.1$  & \textbf{36.983 $\pm$ 5.181}  & 35.167 $\pm$ 4.419  & 36.688 $\pm$ 2.521  & 36.525 $\pm$ 0.754 \\

\textbf{PhysioMTL (kernel)}       & $\alpha=0.1, l_k=20$  & 37.343 $\pm$ 5.173  & \textbf{ 34.379 $\pm$ 4.637}  & \textbf{36.234 $\pm$ 2.298}  & \textbf{35.677 $\pm$ 1.522} \\

\bottomrule
\end{tabular}}
\vspace{-10pt}
\caption{Out-of-sample HRV prediction results on the AHMS dataset, evaluated using RMSE. We randomly choose from a given percentage \{$80\%$, $60\%$, $40\%$, $20\%$\} of subjects for training and then evaluate the model on the remaining held-out test subjects. Best performance for each percentage of training data is bolded.}
\label{table:AHMS}
\end{table*}

\begin{figure*}[h!]
\floatconts
  {fig:ahms_kernel}
  {\vspace{-20pt}
  \caption{The counterfactual analysis on the AHMS dataset. We vary the attributes of a hypothetical subjective's task-wise features and investigate the resulting HRV variational. Each feature investigated (age, BMI, activity, sleep, and VO2\textsubscript{max}) is continuous, however, for ease of visualization we vary each feature on a discrete grid.
  }}
  {%
    \centering
    \subfigure[Age]{\label{fig:ahms_hrv_age_kernel_2}%
      \includegraphics[width=0.29\linewidth]{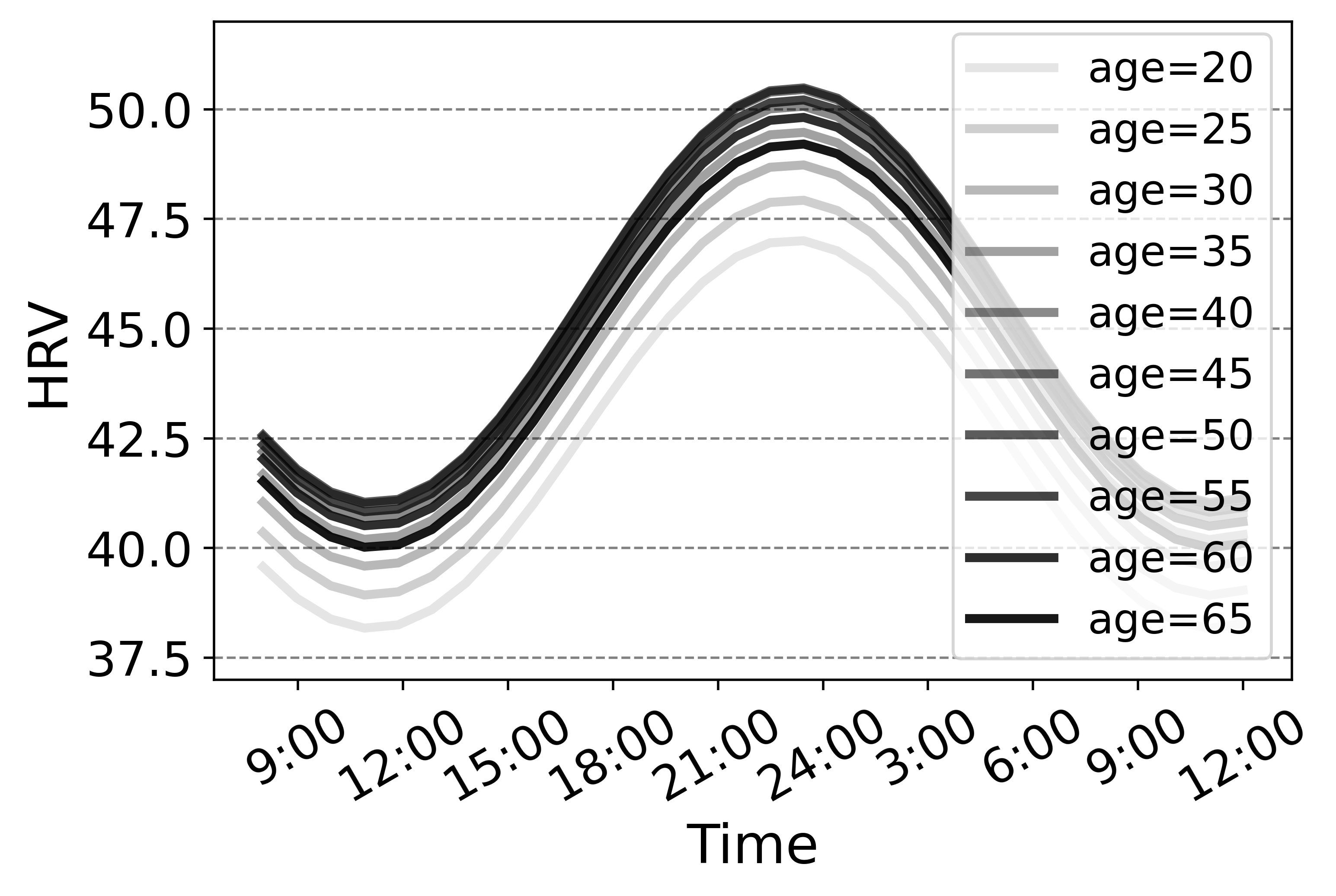}}\hfill%
    \subfigure[BMI]{\label{fig:ahms_hrv_bmi_kernel_2}%
      \includegraphics[width=0.29\linewidth]{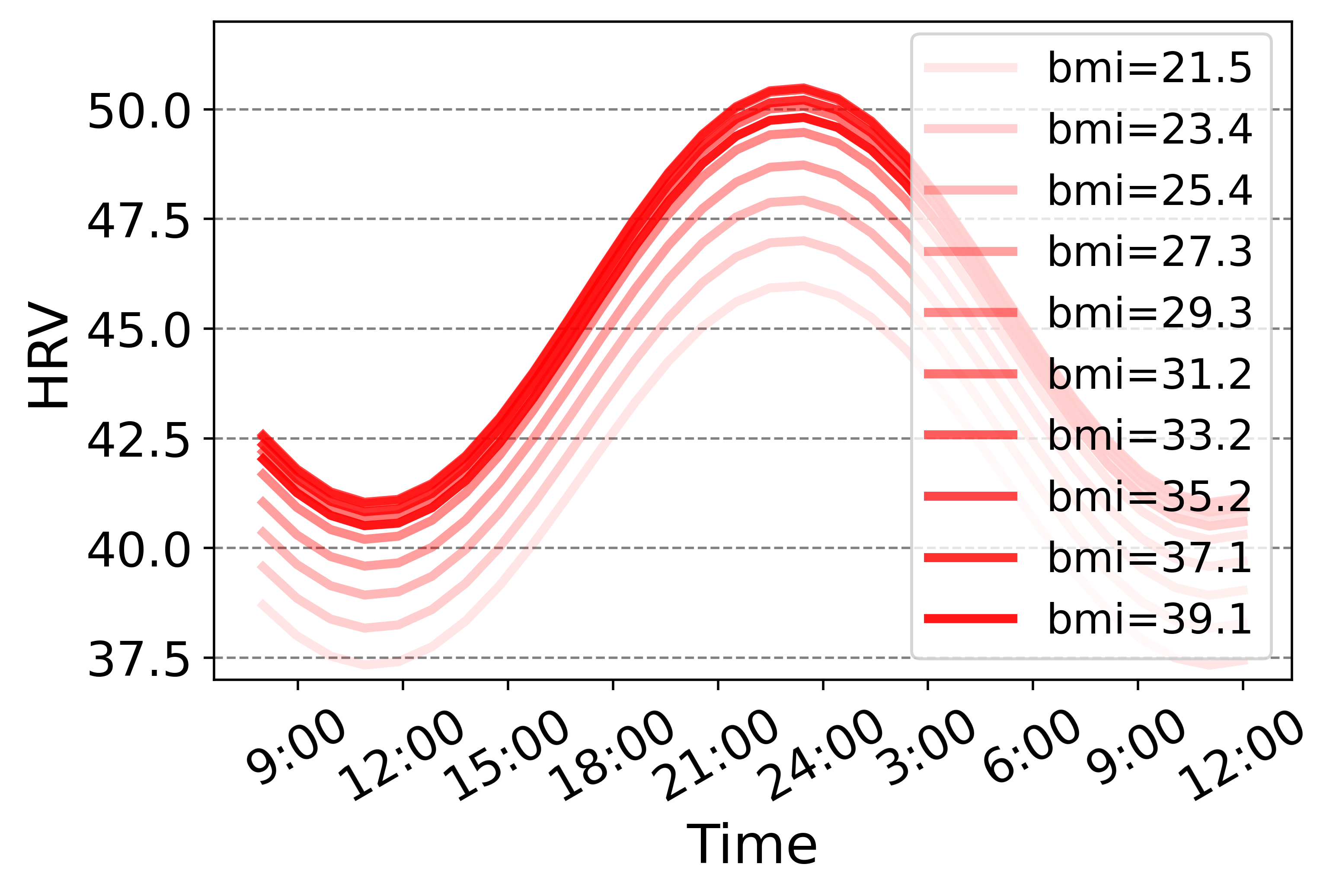}}\hfill
    \subfigure[Activity]{\label{fig:ahms_hrv_activity_kernel_2}%
      \includegraphics[width=0.29\linewidth]{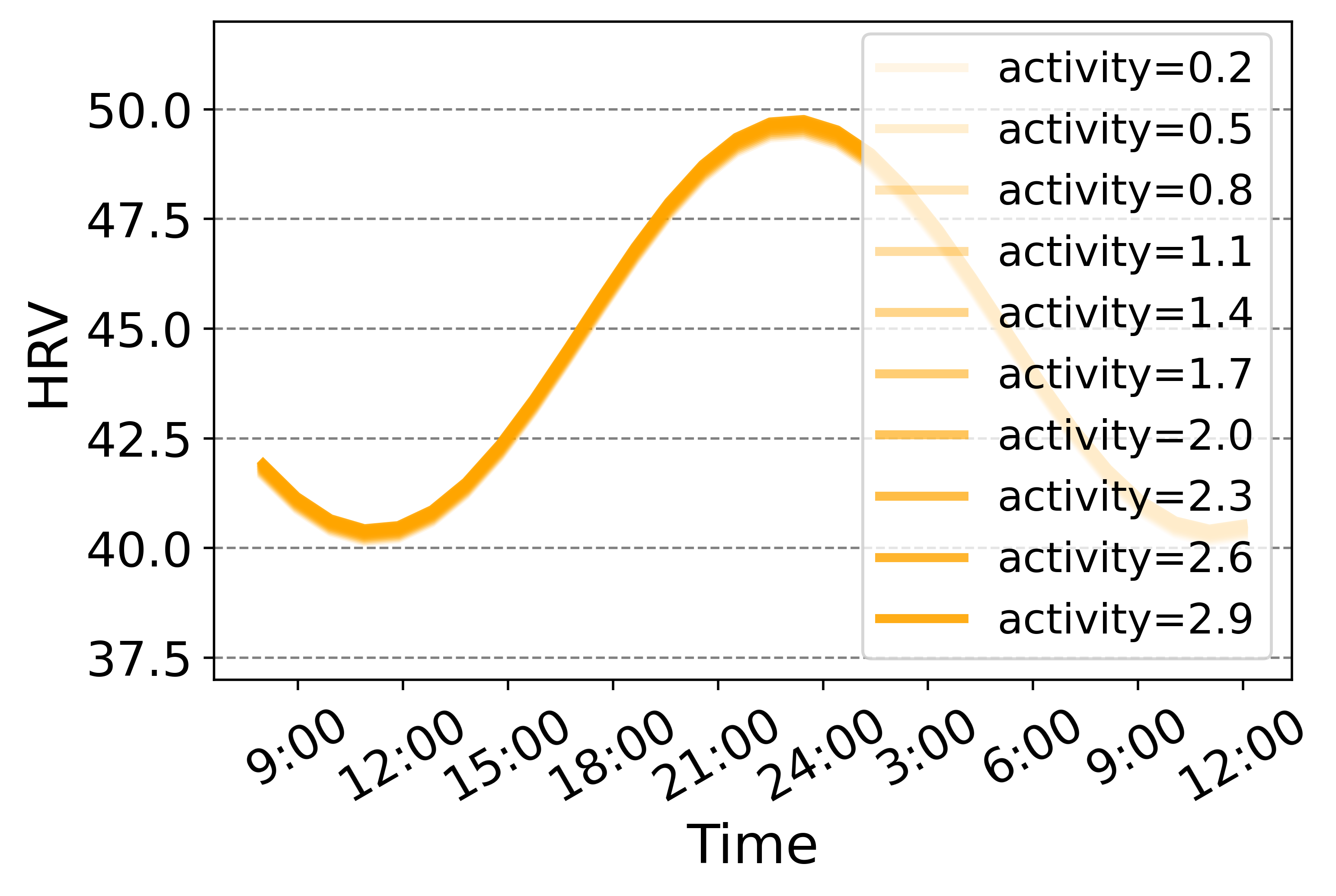}}\hfill%
    \newline
    \centering
    \subfigure[Sleep]{\label{fig:ahms_hrv_sleep_kernel_2}%
      \includegraphics[width=0.29\linewidth]{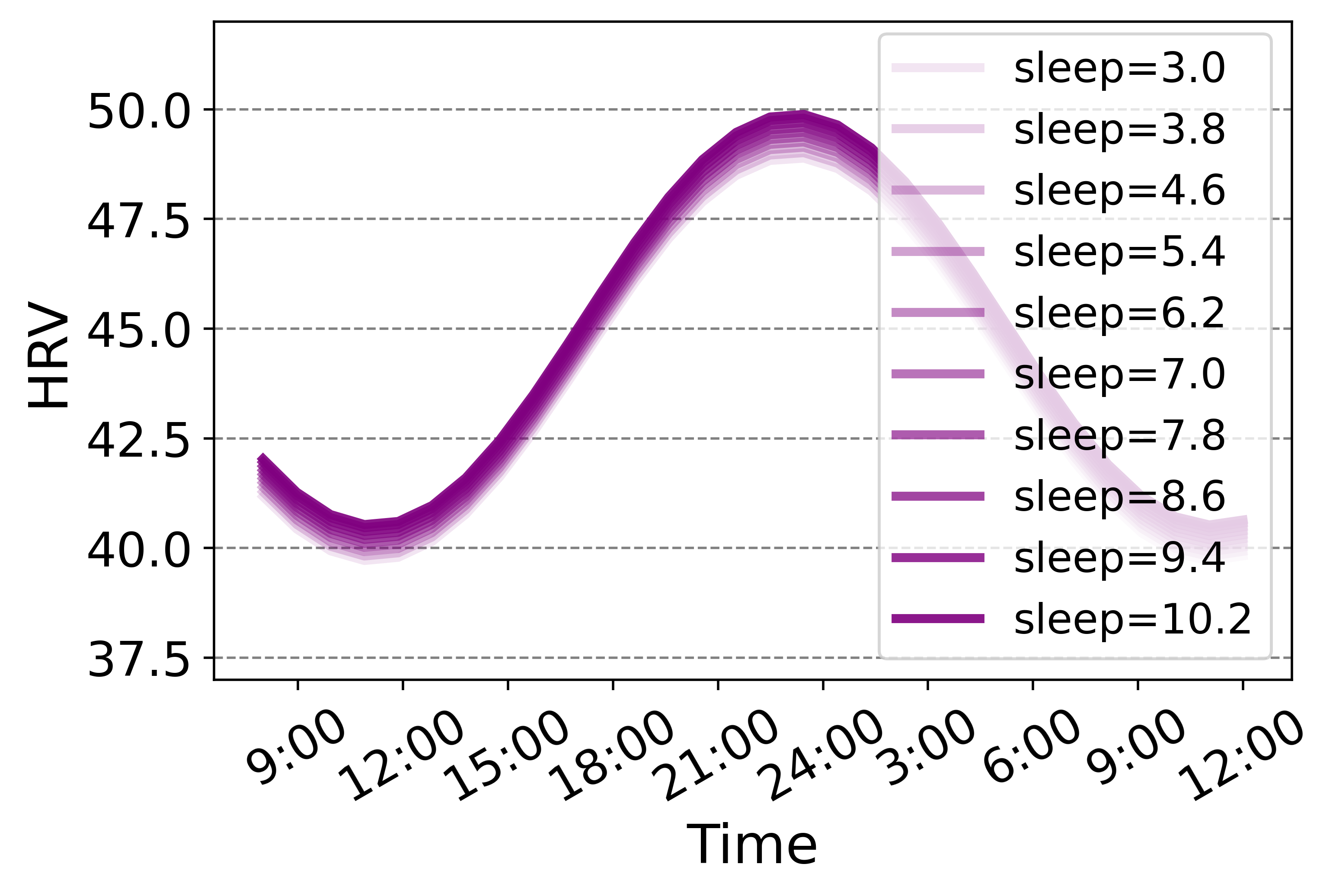}}\hfill%
    \subfigure[VO2\textsubscript{max}]{\label{fig:ahms_hrv_stress_kernel_2}%
      \includegraphics[width=0.29\linewidth]{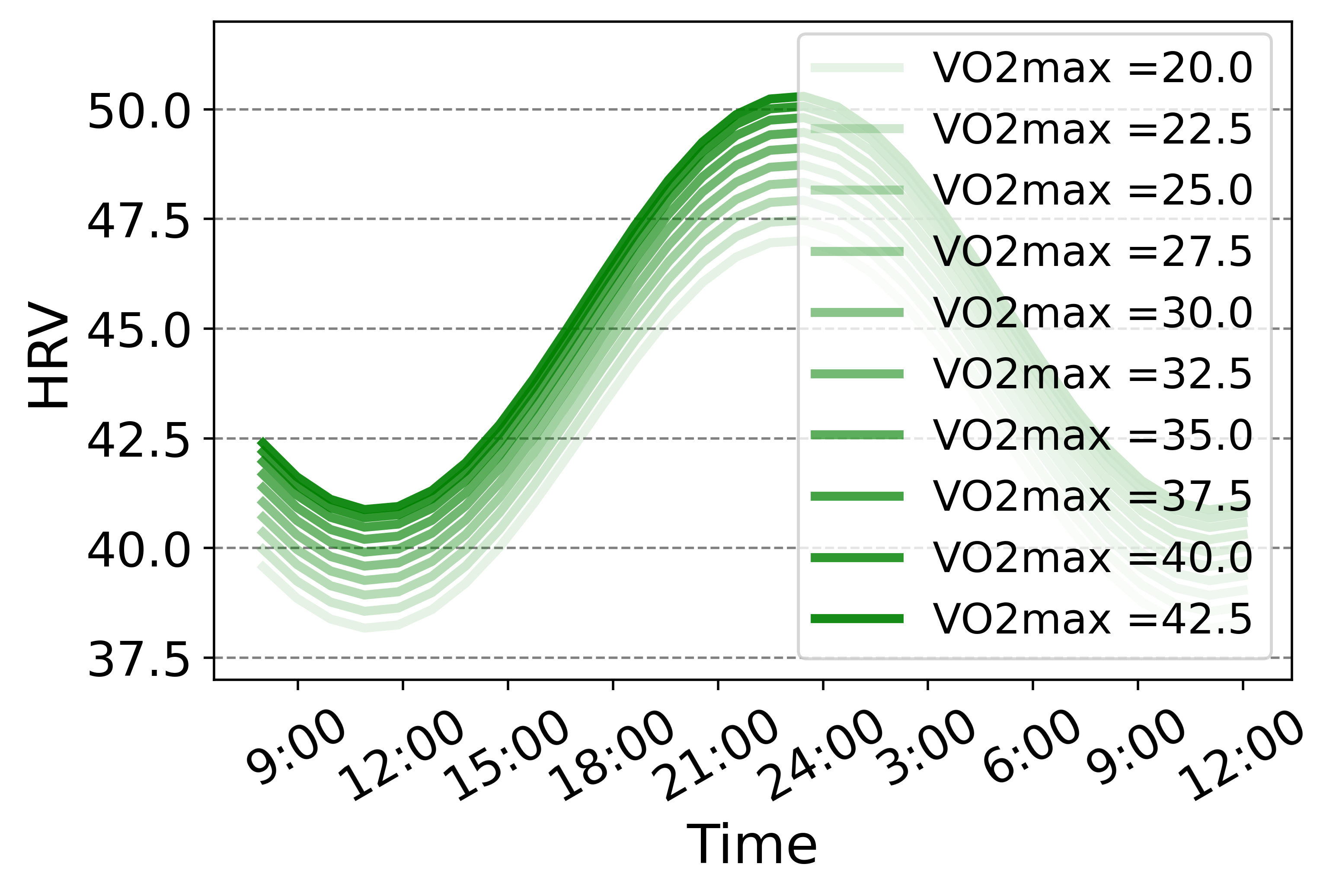}}\hfill
    \subfigure[Baseline]{\label{fig:ahms_hrv_sex_kernel}%
      \includegraphics[width=0.29\linewidth]{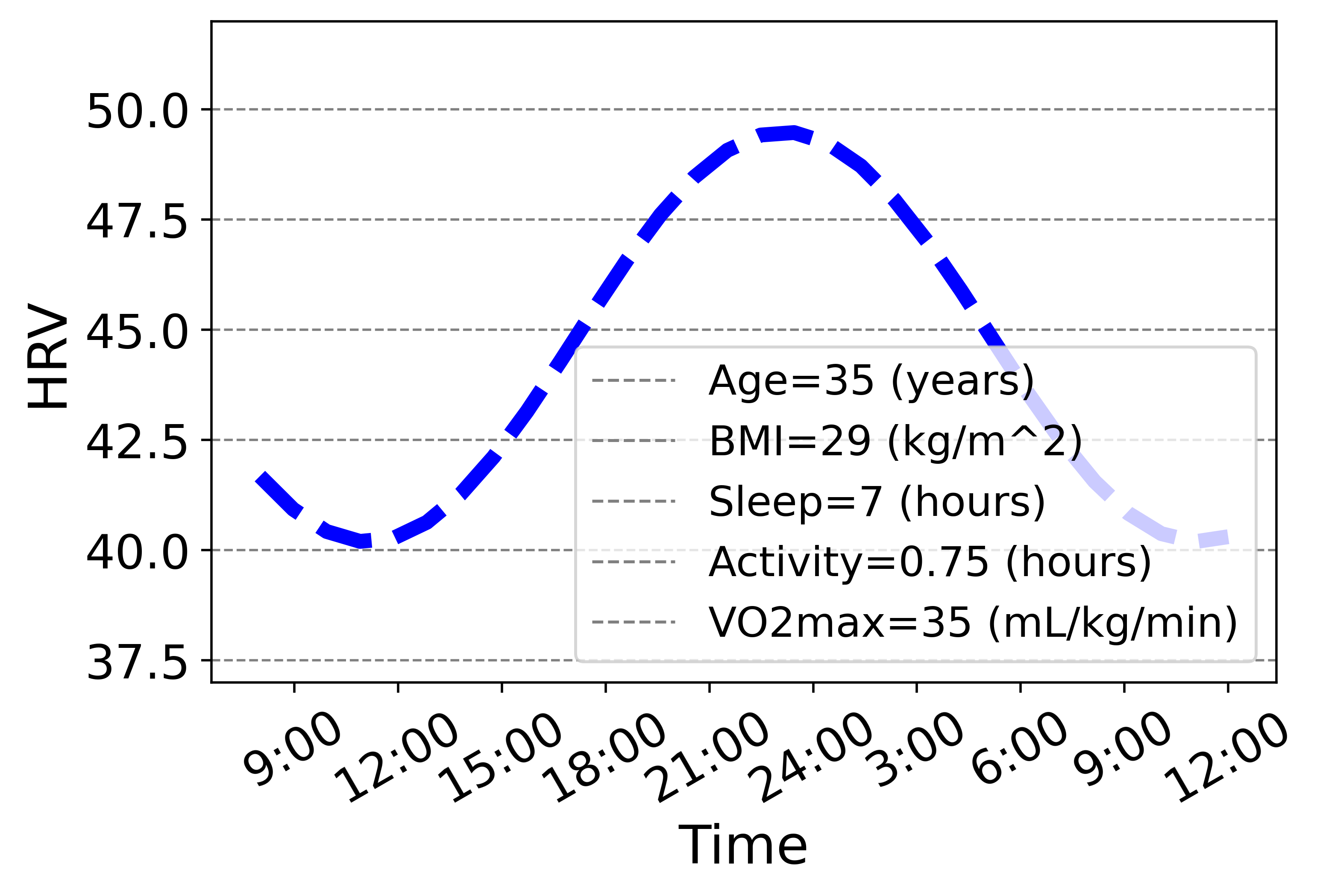}}\hfill%
  } 
\end{figure*}

\begin{table*}[h!]
\centering 
\resizebox{\textwidth}{!}{
\begin{tabular}{lrrrrrr} 
\toprule
Method  & hyperparameter & $80\%$ & $60\%$ & $40\%$ & $20\%$ &  \\
\midrule

Global average   & n/a  & 26.422 $\pm$ 3.490  & 26.622 $\pm$ 1.220 & 27.043 $\pm$ 1.613 & 26.998 $\pm$ 1.073   \\
 
Single Lasso  & $\alpha=0.9$    & 28.950 $\pm$ 3.740  & 28.498 $\pm$ 1.550 & 29.412 $\pm$ 1.657 & 28.812 $\pm$ 1.843  \\

Multi-task Group Lasso   & $\alpha=0.9$   & 28.816 $\pm$ 3.740  & 28.490 $\pm$ 1.529 & 29.177 $\pm$ 1.684 & 28.537 $\pm$ 1.817  \\

Multi-level Lasso   & $\alpha=0.9$   & 28.927 $\pm$ 3.728  & 28.627 $\pm$ 1.555 & 29.358 $\pm$ 1.664 & 28.758 $\pm$ 1.884  \\

Dirty model     & $\alpha=0.9$   & 28.816 $\pm$ 3.740  & 28.490 $\pm$ 1.528 & 29.176 $\pm$ 1.684 & 28.537 $\pm$ 1.816  \\
                    
MTW      & $\alpha=0.9$   & 28.130 $\pm$ 3.802  & 27.689 $\pm$ 1.503 & 28.422 $\pm$ 1.832 & 27.953 $\pm$ 1.711   \\
ReMTW  & $\alpha=0.9$  & 28.175 $\pm$ 3.803  & 27.730 $\pm$ 1.508 & 28.464 $\pm$ 1.824 & 27.995 $\pm$ 1.716  \\
                    
\textbf{PhysioMTL (linear)}     & $\alpha=0.1$  & 27.787 $\pm$ 3.066  & 28.358 $\pm$ 1.252  & {30.124 $\pm$ 1.644}  & {33.453 $\pm$ 4.298} \\
\textbf{PhysioMTL (kernel)}     & $\alpha=0.1$, $l_k=20$  & \textbf{26.338 $\pm$ 3.371}  & \textbf{ 26.470 $\pm$ 1.247}  & \textbf{26.965 $\pm$ 1.660}  & \textbf{26.875 $\pm$ 0.923} \\

\bottomrule
\end{tabular}
}
\vspace{-10pt} 
\caption{Out-of-sample HRV prediction results on the MMASH dataset, evaluated using RMSE. We randomly choose from a given percentage \{$80\%$, $60\%$, $40\%$, $20\%$\} of subjects for training and then evaluate the model on the remaining held-out test subjects. Best performance for each percentage of training data is bolded.}
\label{table:MMASH}
\end{table*}

\begin{figure*}[h!]
\floatconts
  {fig:mmash_kernel}
  {\vspace{-20pt}  
  \caption{The counterfactual analysis on the MMASH dataset. We vary the attributes of a hypothetical subjective's task-wise features and investigate the resulting HRV variational. Each feature investigated (age, BMI, activity, sleep, and stress) is continuous, however, for ease of visualization we vary each feature on a discrete grid. 
  }}
  {%
    \centering
    \subfigure[Age]{\label{fig:mmash_hrv_age_kernel_2}%
      \includegraphics[width=0.29\linewidth]{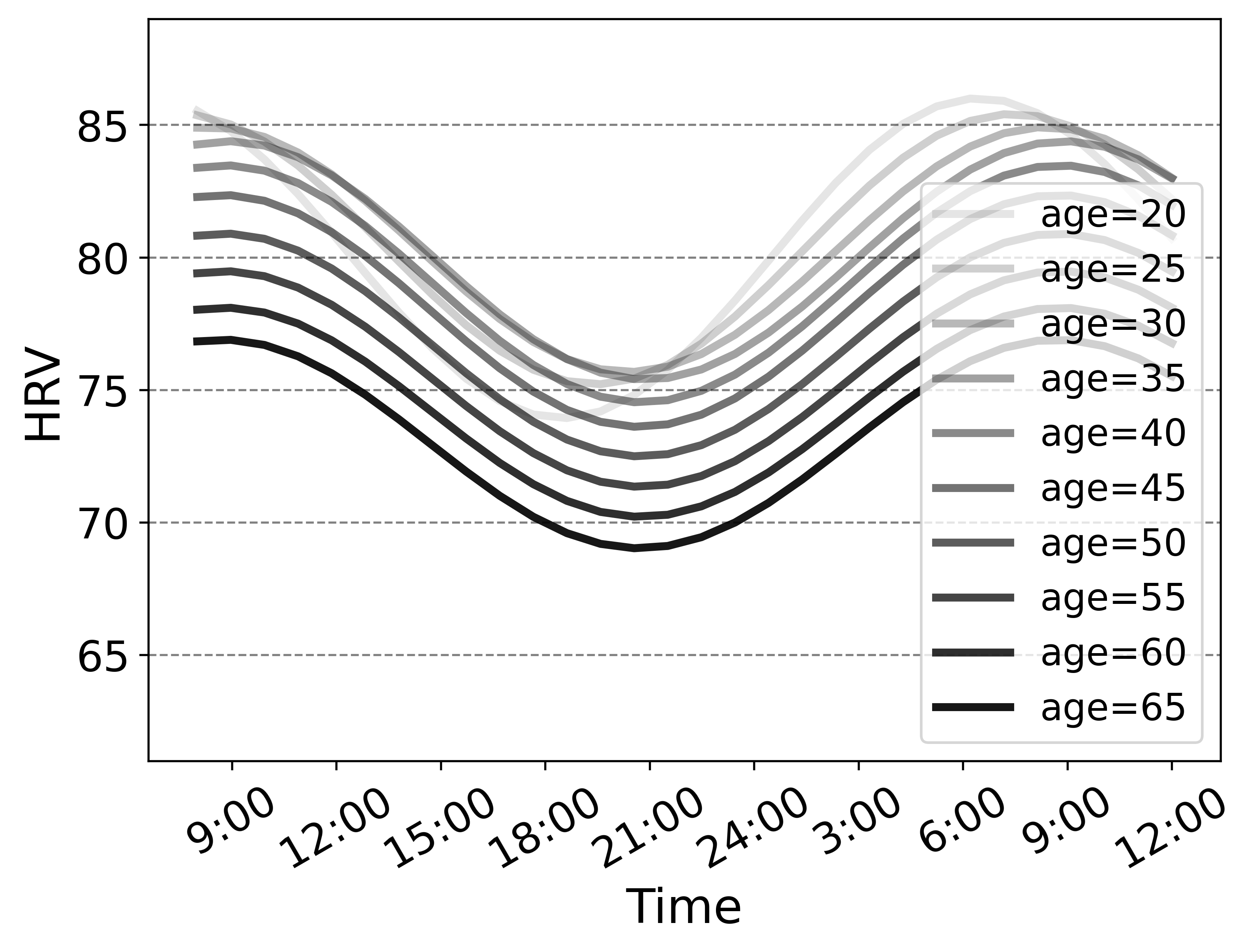}}\hfill%
    \subfigure[BMI]{\label{fig:mmash_hrv_bmi_kernel_2}%
      \includegraphics[width=0.29\linewidth]{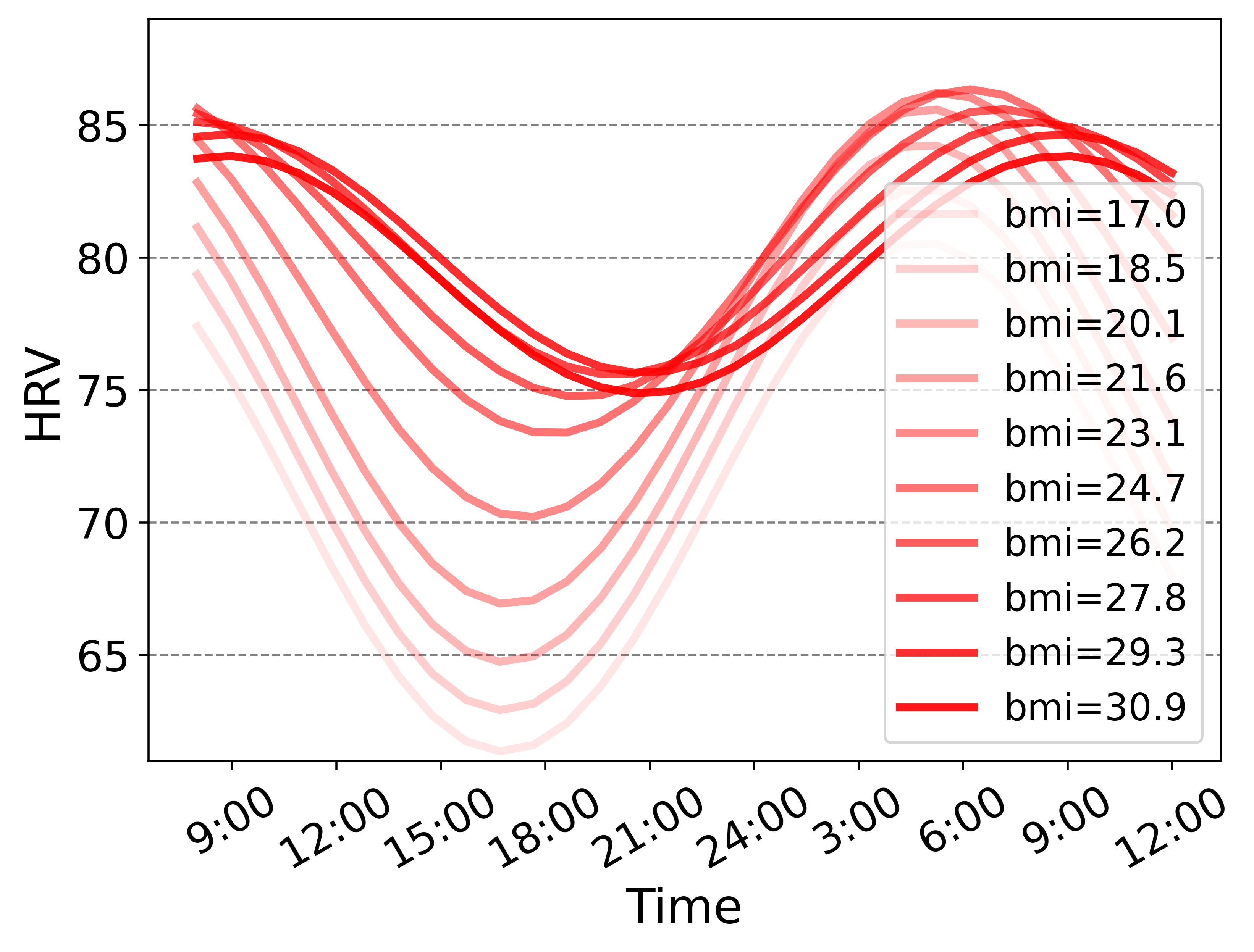}}\hfill
    \subfigure[Activity]{\label{mmash_hrv_activity_kernel_2}%
      \includegraphics[width=0.29\linewidth]{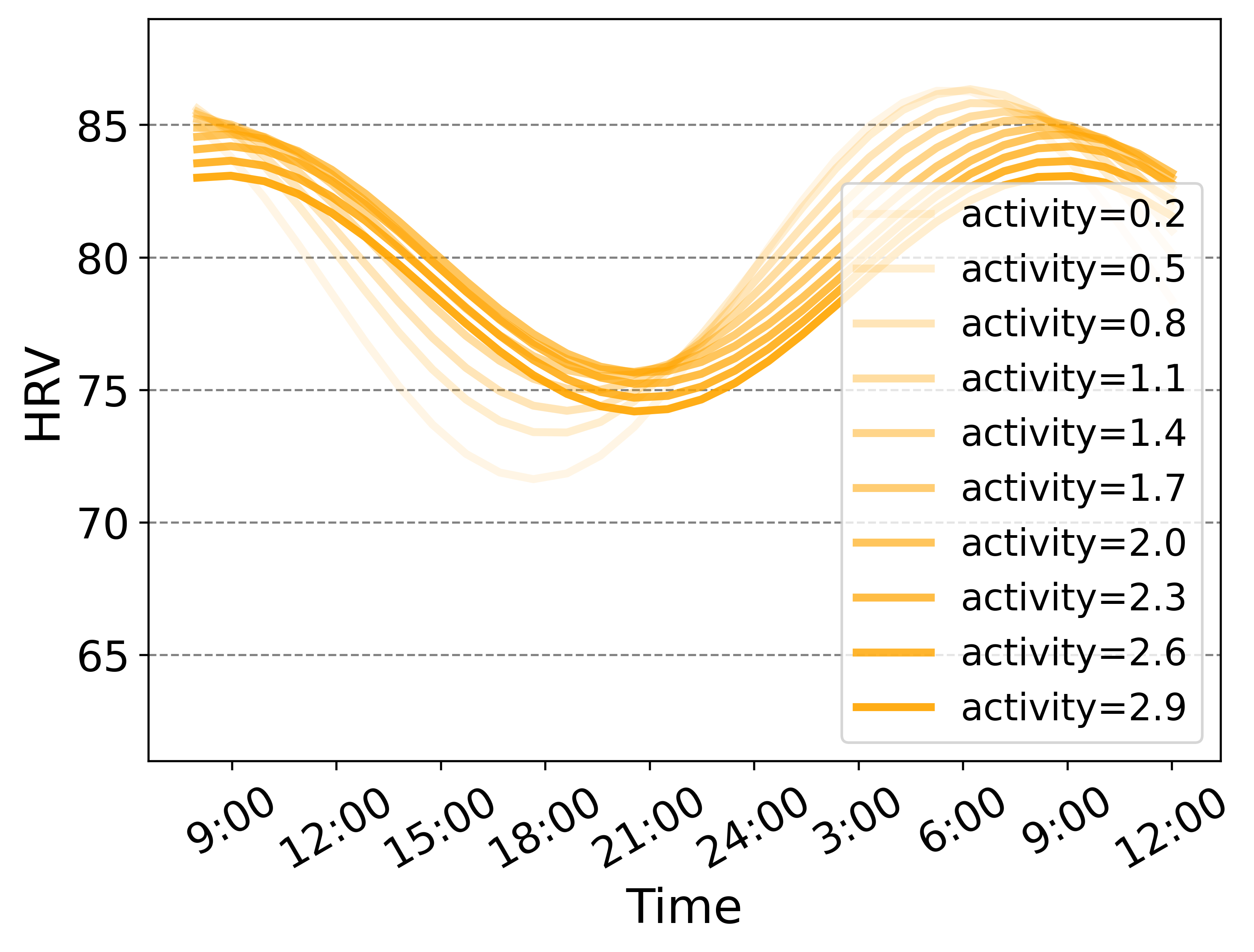}}\hfill%
    \newline
    \centering
    \subfigure[Sleep]{\label{mmash_hrv_sleep_kernel_2}%
      \includegraphics[width=0.29\linewidth]{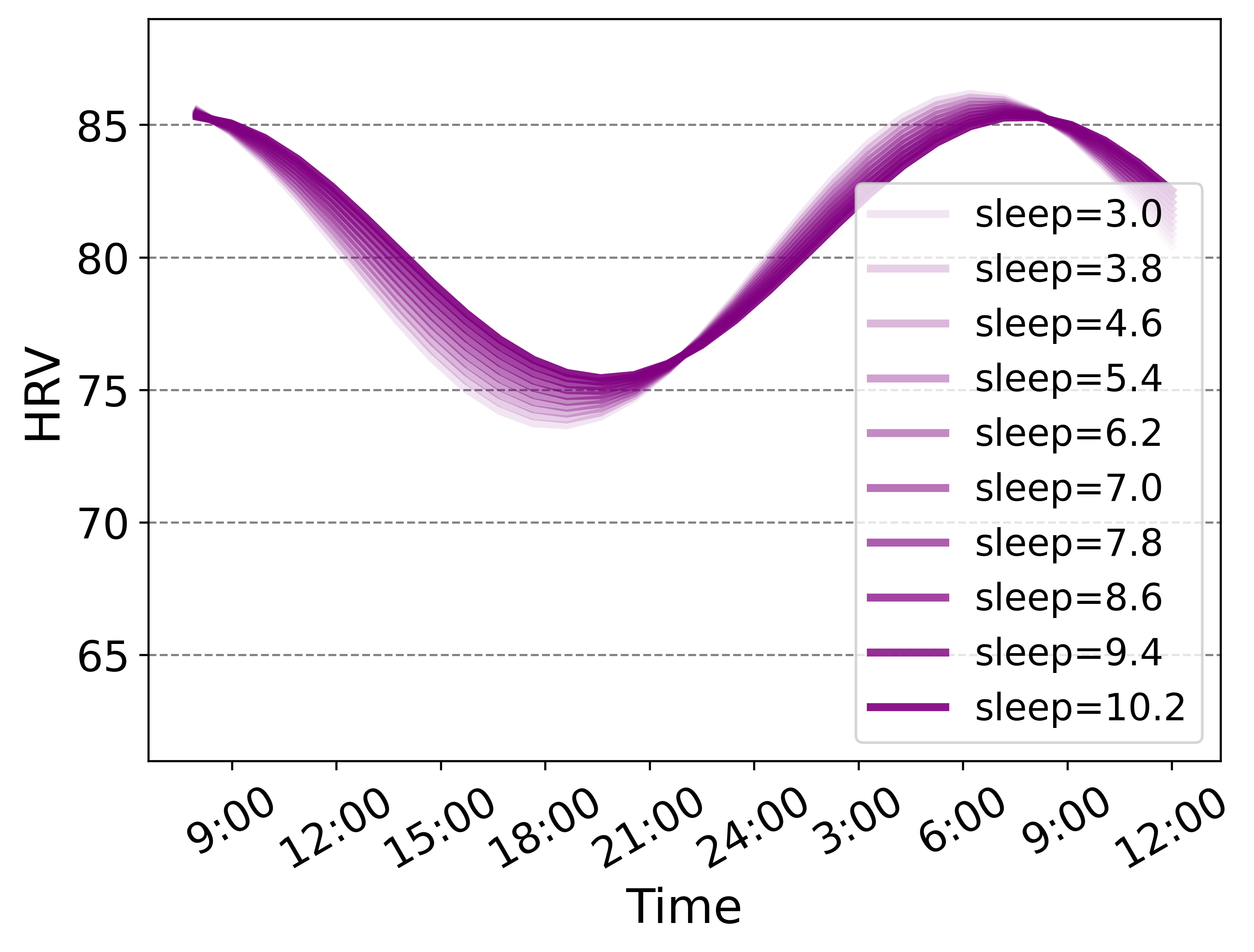}}\hfill%
    \subfigure[Stress]{\label{mmash_hrv_stress_kernel_2}%
      \includegraphics[width=0.29\linewidth]{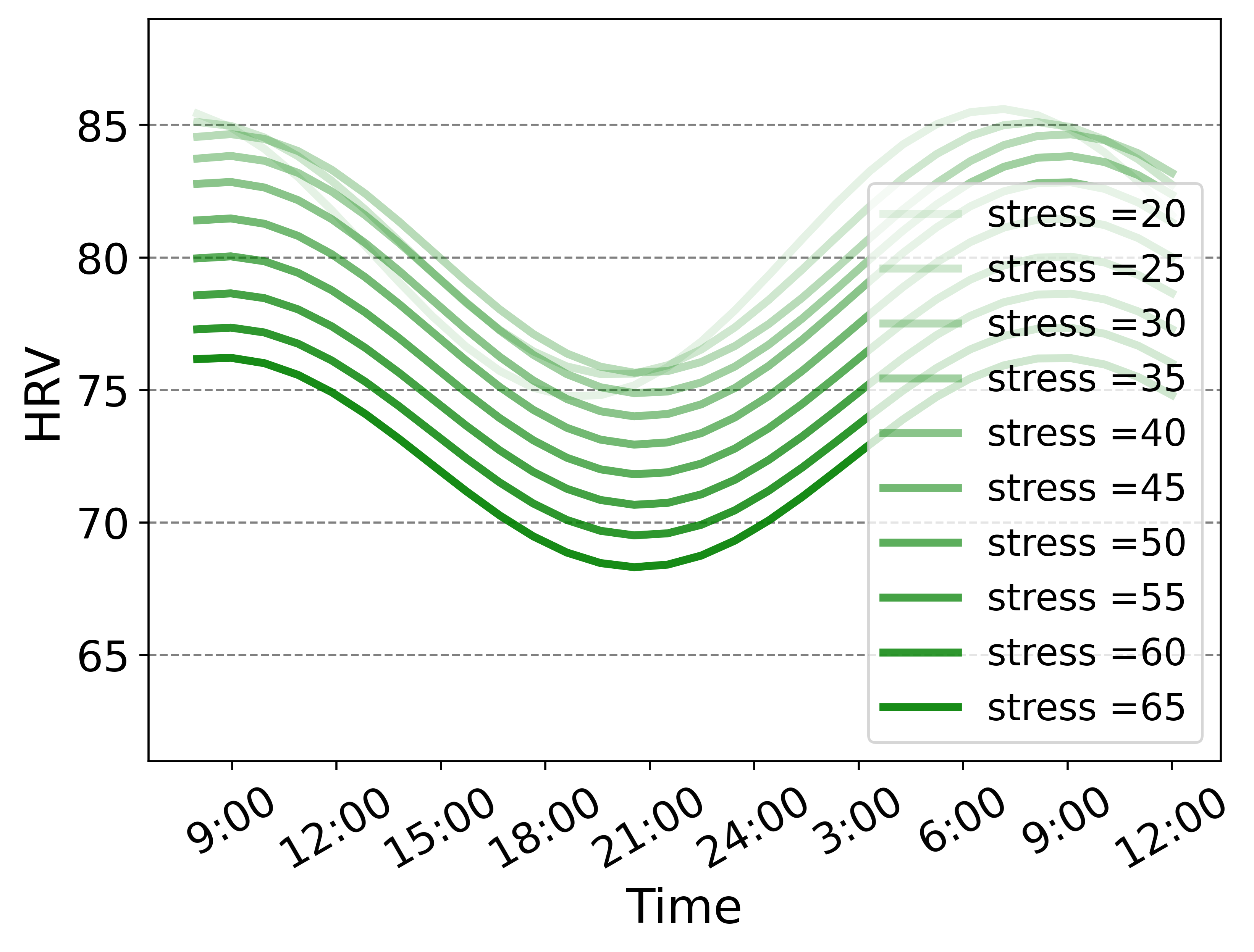}}\hfill
    \subfigure[Baseline]{\label{fig:mmash_hrv_sex_kernel}%
      \includegraphics[width=0.29\linewidth]{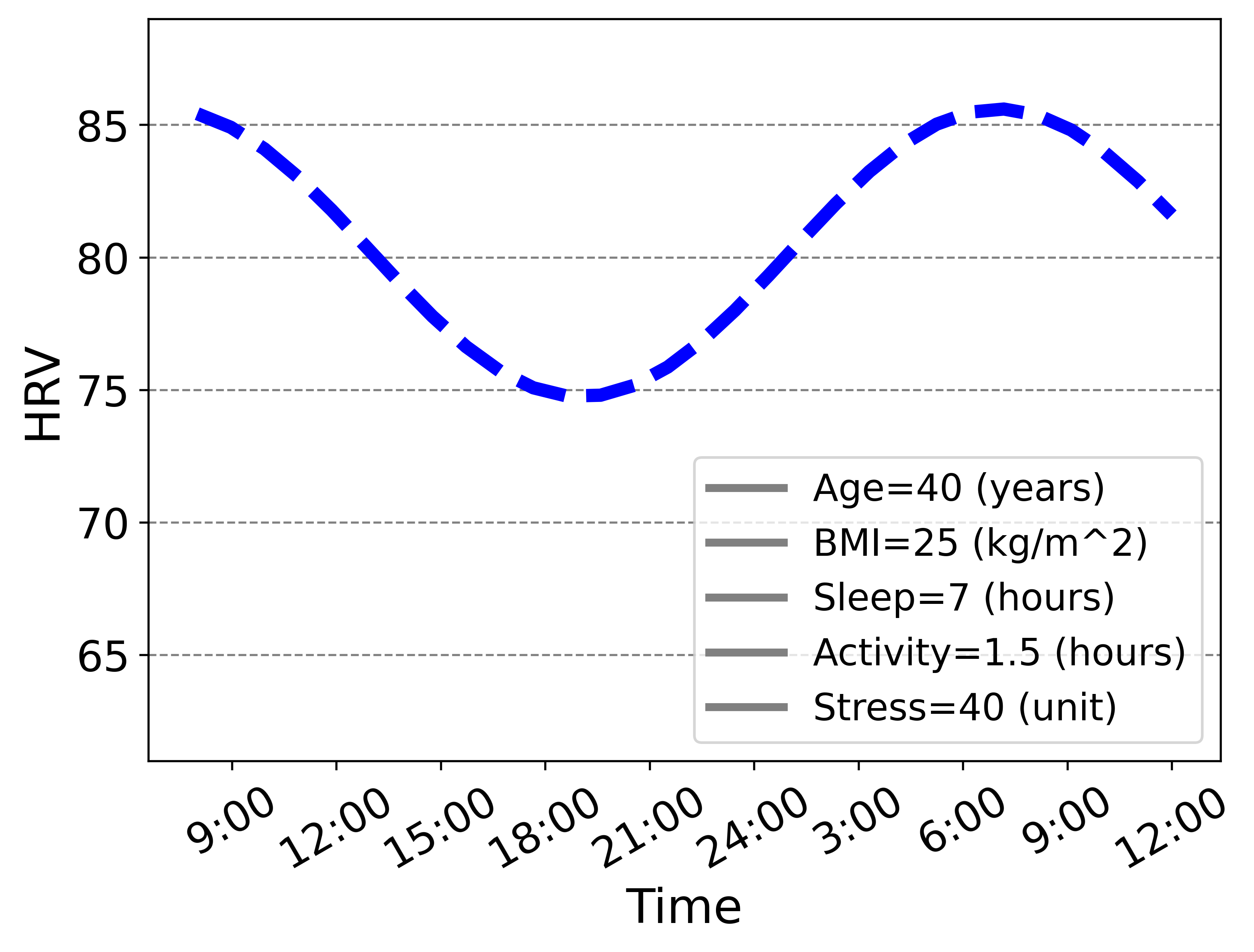}}\hfill%
    
  } 
\end{figure*}

\subsection{Multi-task regression with synthetic data}

We first sample the task-wise features via $s_{t,i} \sim U[p, q]$  where $p$, $q \in \mathbb{R}$ are hyperparameters that control the task distribution.
Then we generate the feature-label pairs of each task via the following equation
\begin{equation}
\label{eq:toy_example}
    y_{t,i} = m(s_t) + A(s_t) \sin(\frac{2 \pi}{T} \tau_{t,i} + \phi(s_t) ) + e_{t,i}
\end{equation}
where $m(s_t)$, $A(s_t)$ and $\phi(s_t)$ are the groundtruth parameters for each task that are controlled by the task-wise feature $s_t$ by $m = a_m + b_m \times s_t$, $A = a_A + b_A \times s_t$ and $\phi = a_\phi + b_\phi \times s_t$. Here $e_{t,i} \sim N(0, \sigma_t^2)$ is the task-specific noise. 
Then we randomly split the data into training and testing tasks, $\{s_{t,i}\}_i^{n_t}$ and $\{s_{e,i}\}_i^{n_e}$, where $n_t$ and $n_e$ are the number of training and testing tasks respectively.  To measure the similarity between the training and testing tasks, we use a 2-Wasserstein distance of uniform empirical measures $W(\mu(s_t), \nu(s_e))$, where a larger Wasserstein distance indicates a larger divergence between the training and testing tasks. 

We compare the performance of our method against \textit{Group Lasso}~\citep{yuan2006model_group_lasso}, \textit{Multi-level Lasso}~\citep{lozano2012multi_level_lasso}, \textit{Dirty models}~\citep{jalali2010dirty}, \textit{Multi-task Wasserstein} and \textit{Reweighted Multi-task Wasserstein}~\citep{janati2019wasserstein_mtl}.
The prediction error of our method is lowest amongst all methods we compared across a range of divergences between training and testing data (Figure~\ref{fig:toy_example_quant}).
Additionally, the prediction error of other methods increases as the divergence between the training and testing task becomes larger, however, our model error increases at a far slower rate. This experiment validates that PhysioMTL is generalizable to out-of-sample tasks. In addition, in Figure \ref{fig:toy_example_uneven} we demonstrate that PhysioMTL exploits the information shared among tasks and performs well on a given task that has samples constrained to a limited portion of time. Note that this is a quite common case in practice for wearable devices.

\subsection{ Experiments on real-world datasets} 

We conducted experiments on two real-world datasets: (1) the \textit{Apple Heart and Movement Study (AHMS)} dataset and (2) the \textit{Multilevel Monitoring of Activity and Sleep in Health people (MMASH)} dataset.

\subsubsection{The AHMS dataset}
The AHMS protocol allows participants to collect a comprehensive set of biometrics through wearable devices, thus providing a suitable dataset for studying the assessment of physiology status through wearable devices.
In the AHMS dataset, \textit{HRV} values are obtained by calculating the root mean square of successive differences between beat-to-beat measurements (RMSSD) captured by a heart rate sensor. Heart rate measurements are automatically recorded throughout the day while participants wear an activity tracker, thus there are multiple measurements each day. We obtain the task-wise feature vector for each subject as follows: the anthropometric characteristics including \textit{age}, \textit{biological sex}, \textit{height}, and \textit{weight} are self-reported by study participants.
We use \textit{time in bed} to indicate the sleep quality.
We use hours of exercise to serve as an indicator of physical activity.
We select a random set of subjects ($100$) that are split evenly between two self-reported genders (Female/Male), and have a minimum of 50 HRV measurements. We consolidate HRV measurements over a five month period of time because of the lack of measurement density within each 24 hour period, and average the demographic and lifestyle factors over that period if there are any changes.

\subsubsection{The MMASH dataset}
In the Multilevel Monitoring of Activity and Sleep in Healthy people (MMASH) dataset ~\citep{rossi2020public_mmash},  
$22$ healthy male subjects provided a set of 24 hours of continuous measurements including inter-beat intervals data, wrist accelerometry data, sleep time and quality data, physical activity data, psychological characteristics (e.g. anxiety status, stressful events, and emotion declaration).  

Since HRV measurements are not directly provided, we first subdivide the inter-beat interval data into separate 5 minute intervals, and then we compute the RMSSD \textit{HRV}~\citep{shaffer2017overview_hrv_measure}, which is the root mean square of successive differences between normal heartbeats, within each 5-minute interval. The length of interval is chosen as 5 minutes since it is the conventional short-time recording standard.
We use the following attributes as task-wise features: \textit{Activity, Sleep, and Stress}.
Individual anthropometric characteristics such as \textit{Age, Height, and Weight} are provided as well.
We use \textit{total minutes in bed} to denote the sleeping information and convert it to \textit{total hours in bed} for consistency. The MMASH dataset provides an activity diary for each subject, we use the total hours of \textit{medium (e.g. fast walking and cycling)} and \textit{heavy (e.g. gym, running)} to indicate physical activity. Finally, each subject provides a subjective measure of stress denoted as the Daily Stress Inventory (DSI) score. 
We process the data with the following steps: (1) For each subject, we remove the outliers with z-score greater than $2.5$ after computing RMSSD HRV values from 5-minute intervals, (2) we exclude the subjects that have particularly abnormal measurements (i.e., subject 4 has an average RMSSD of $318$), and (3) since subjects 11 and 18 do not provide sleep data and age, respectively, these missing values are imputed with data-wide means. Our pre-processing led us to consider 21 out of 22 subjects.

\subsection{Experimental procedure}
To investigate the generalization, we evaluate the model on those tasks that are \textit{unobserved} during training. We randomly split the tasks into training and testing task sets, and then evaluate the performance of each method using root mean square error (RMSE) of prediction. We repeat the selection-evaluation process for $n=10$ times and report the mean and standard deviation of each method's RMSE. We select the $k$-nearest tasks for baseline MTL methods when predicting \textit{out-of-sample} tasks. To enable a meaningful comparison, we use the same task similarity metric $C_s(\cdot, \cdot)$ for all MTL methods where we encourage all the attributes in the task-wise feature vector to have an equal influence on task similarity.

For experiments on the AHMS dataset, we use a global mean estimator and Lasso estimator independently run on each task as a baseline. We additionally compare Multi-task Lasso~\citep{obozinski2006multi_lasso} and Multi-task Elastic-Net~\citep{zou2005regularization_elastic} with our approach. For the MMASH dataset, we compare our method with~\textit{Group Lasso}~\citep{yuan2006model_group_lasso}, \textit{Multi-level Lasso}~\citep{lozano2012multi_level_lasso}, \textit{Dirty models}~\citep{jalali2010dirty}, \textit{Multi-task Wasserstein} and \textit{Reweighted Multi-task Wasserstein}~\citep{janati2019wasserstein_mtl}. We select the best hyperparameter $\alpha$ for all baseline methods.
  
\subsection{Results and physiological patterns}

\subsubsection{Predictive performance}
Our method has superior performance on both datasets compared with all baseline methodologies. The quantitative results on the AHMS dataset are given in Table~(\ref{table:AHMS}). 
In Table~(\ref{table:MMASH}), we provide a comprehensive comparison on the MMASH dataset. We test the extreme case where only ($20\%$) of tasks are available for training and found PhysioMTL with a nonlinear optimal transport map achieves the best performance. While PhysioMTL with a linear map is competitive across many scenarios, with a nonlinear transport map PhysioMTL has lower error than all other methods. It is worth mentioning that HRV measurements are extremely noisy and there exist numerous confounding variables. Therefore, even a small improvement of RMSE means a more accurate estimation of HRV circadian rhythm parameters. A more illustrative example is given in Figure (\ref{fig:toy_example_uneven}), where the RMSE errors of all methods are similar, but the circadian rhythm estimated by the PhysioMTL is closer to the groundtruth.

\subsubsection{Counterfactual analysis 
}
The physiological constraints we encode in PhysioMTL allow us to investigate the affect of acute and chronic stressors on \textit{cardiovascular activity} assessed through HRV. We can construct counterfactuals by setting a baseline task feature vector and varying one dimension at a time while holding the remaining dimensions constant.

In Figure~(\ref{fig:mmash_kernel}), we display a set of HRV variations across several dimensions. In general, all the results are consistent with existing physiological studies. \textit{Age} is a dominant factor, as aging in healthy subjects is typically associated with a decline in HRV. Our results also show aging will lead to a decrease in the circadian rhythm amplitude. We found \textit{BMI} has a non-monotonic relationship to HRV mean value.
Also, an increasing \textit{sleep hours} will lead to a higher and more stable circadian rhythm. Interestingly, we observed the HRV pattern first goes up as \textit{activity} increases and then decreases. This trend reflects the phenomenon that light endurance training increases acute HRV but higher exertion bouts can cause decreases in HRV~\citep{plews2014heart}. Finally, our model shows stress and anxiety present as lower values of HRV.

\section{Conclusion}
We investigate how acute and chronic factors affect HRV rhythms.
We leverage established sinusoidal analysis of the ANS system and optimal transport theory to develop the PhysioMTL model. The tasks are viewed from a geometric and probabilistic perspective, and an optimal transport map captures a general correspondence between tasks that are similar and generalizes to unseen samples. We validate our method on synthetic and real-world physiological datasets and show our approach outperforms existing MTL methods. Our model learns an approximation of the true latent HRV pattern for each individual and serves as an accurate prediction for new individuals based on a few pieces of easily captured information. 

Beyond prediction, our method lends itself to analyzing counterfactual scenarios. This allows us to investigate the hypothetical question of what underlying physiology would have presented as given an unobserved setting of features. We use this counterfactual to show the effect of our captured demographic and lifestyle covariates on HRV rhythms.

We note that we do not directly compare to meta-learning and transfer learning methods since we evaluate all methods on completely unseen tasks and do not consider a fine-tuning step on prediction tasks. However, the predicted function (model parameters) can actually serve as a prior initialization for an unseen task and enable fine-tuning on the target domain. Thus, our method can be considered a potentially general meta-learning formulation where the optimal transport map is a meta-learner. Towards this goal, future work will also consider extending our framework to more flexible and expressive models, such as function spaces modeled by deep neural networks. In this case, it may be possible to directly predict the parameters or learn suitable priors over the parameters of a task-specific neural network. We also consider the possibility of learning a larger network with a shared subcomponent that performs representation learning over tasks, and a task-specific subcomponent with priors learned from training data.

Finally, we study a single physiological indicator, HRV, since the diurnal patterns that serve as the baseline for our method are well-studied.  The covariates we study have also been analyzed to some degree previously, thus there is prior evidence to support our findings. However, our framework is quite general. By changing the form of the individual task regressor and the task relatedness kernel, our methodology could be used to investigate a number of other physiological and non-physiological signals where captured covariates are thought to explain a fair amount of variation in the signal. A further extension to our work could jointly model multiple signals in order to share predictive power captured in the covariates across multiple associated signals.

\section*{Institutional Review Board (IRB)}
The Apple Heart and Movement study was conducted in collaboration between Apple and Brigham and Women's hospital, the data were collected in accordance with an Advarra IRB approved study protocol.

\acks{We would like to thank Guillermo Sapiro, Calum MacRae, Nick Foti, Russ Webb, Haraldur Hallgrímsson, Gary Shin, Laixi Shi, and Jielin Qiu for valuable feedback and comments on our manuscript.}

\bibliography{main}

\newpage

\onecolumn

\appendix

\section{Further Literature  Review}\label{apd:first}

\subsection{Machine learning for healthcare}

The Spatio-temporal group lasso (STGL-MTL) \citep{romeo2020novel_spatiotemp_ijcai_health} encodes the task relatedness with a regularization term and a graph structure for diabetes prediction. 
Within the MTL framework, the combination of task labels can provide auxiliary tasks \citep{hsieh2021boosting_combine_aux} and aid in ECG phenotyping prediction. 
A temporal asymmetric MTL model \citep{nguyen2020clinical_mtl_symm} is proposed to enforce knowledge transfer only between relevant tasks to approach safety-critical scenarios such as clinical risk prediction. 
A multi-task generative model \citep{wang2020graph_mtl_vae} has also been used for clinical topic modeling. Multi-view deep learning models are develop to deal with the multimodality of human emotion,  electroencephalography (EEG) and 
electrocardiogram (ECG)~\citep{qiu2018multi_emo,liu2021comparing_multi_emo}.
Recently, Optimal Transport (OT) related methods~\citep{qiu2022optimal_ECG} have improved the performance of deep learning methods for electrocardiogram (ECG).

\subsection{Multi-task learning and task relatedness}

Various multi-task learning approaches have been developed to exploit the correlation among tasks. For instance, task clustering approaches \citep{he2019efficient_ccmtl,zhou2011clustered_mtl_jiayu} assume different tasks can be clustered. Feature sharing \citep{evgeniou2007multi_mtl_feature} and low-rank \citep{zhou2011malsar_jiayu_ye_lowrank} approaches assume all tasks are related. Kernel-based methods \citep{bonilla2007kernel} encode task-specific features to obtain task similarity and therefore enhance the prediction performance.

\subsection{Intuition}

To tackle domain adaptation limitations, we incorporate optimal transport map estimation theory with multi-task learning. Optimal transport (OT)~\citep{villani2009optimal_old_new} has attracted considerable interest in the machine learning community.  
OT provides a meaningful distance between probability measures that do not share the same support; OT also formulates a transport map~\citep{zhu2021functional_fot,perrot2016mapping,alvarez2019towards} that pushes forward one distribution onto another. Also, the addition of entropic regularization~\citep{cuturi2013sinkhorn} yields fast computation, thus enabling scalability of OT to larger problems.

\cite{pop2021assessment_HRV_alcohol,jarczok2019circadian_sinusoi} have shown that those individuals that have similar demographic attributes exhibit a similar HRV pattern. To this end, we regard the observed individuals as empirical distributions
evaluated from an underlying distribution and further assume a smooth transport map from the distribution of tasks to the distribution of predictive functions.

\subsection{OT}

the Monge problem consists in finding a Borel map, $T: \mathcal{X} \mapsto \mathcal{Y}$ between $\mu$ and $\nu$. 
that realizes the infimum
\begin{eqnarray}
\label{eq:Monge}
\inf_T  \int_{\Omega} c(x, T(x)) d \mu(x)  \text{ \smallskip {}{} subject to {}{} \smallskip  }   T_{\#}\mu = \nu,
\end{eqnarray}
where $T_{\#}\mu$ denotes the push forward operator of $\mu$ by $T$.
However, the existence of the optimal Monge map $T$ is not always guaranteed---the Monge problem is non-convex and often unfeasible, for example, when the support $\mu$ and $\nu$ are a different number of Dirac delta functions.

\section{Computational details}\label{apd:computation}

\subsection{Derivation of computation algorithms}

We give a local minima solution for the objective Equation. (\ref{eq:obj_fin}) by alternatively minimizing over W and F with regard the loss function:
\begin{eqnarray}
L(\mathbf{F}, \mathbf{W}) = \frac{1}{2}  \sum_{t=1}^T  \| W^T_t X_t  - y_t\|^2_2  \nonumber  +  \alpha \sum_{i,j=1}^T {\pi^*}_{i,j} \| \mathbf{F} \mathbf{s}_i- W_j \|_2^2.
\end{eqnarray}
The algorithm is described in Alg. (\ref{alg_PhysioMTL}) and we give the explicit calculations below. We will use the linear map in derivation but notice that for both ways of transport map parameterization (linear and kernel), the only difference is the size of the $\mathbf{F}$ matrix.

\textbf{Updating $\mathbf{F}$ with $\mathbf{W}$ fixed:} Here the minimization objective becomes
\begin{equation}
    \hat{\mathbf{F}} = \arg \min_{\mathbf{F} \in \mathrm{R}^{d_\theta \times d_s}} L(\mathbf{F}, \mathbf{W})  = \arg \min_{\mathbf{F} \in \mathrm{R}^{d_\theta \times d_s}} \alpha \sum_{i,j=1}^T {\pi^*}_{i,j} \| \mathbf{F} \mathbf{s}_i - W_j \|_2^2.
\end{equation}
We preform gradient descent method to achieve the minimum, where the gradient is:
\begin{equation}
    \nabla_{\mathbf{F}} L(\mathbf{F}, \mathbf{W}) = 2 \alpha \sum_{i=1}^T \sum_{j=1}^T \pi^*_{i,j} (\mathbf{F} s_i - W_j ) s_i^\top. 
\end{equation}

\textbf{Updating \textbf{W} with \textbf{F} fixed:} Now we want to solve
\begin{equation}
     \hat{\mathbf{W}} = \arg \min_{\mathbf{W} \in \mathrm{R}^{T \times d_w}} L(\mathbf{F}, \mathbf{W}) =  \arg \min_{\mathbf{W} \in \mathrm{R}^{T \times d_w}} \frac{1}{2}  \sum_{t=1}^T  \| W^T_t X_t  - y_t\|^2_2  \nonumber  +  \alpha \sum_{i,j=1}^T {\pi^*}_{i,j} \| \mathbf{F} \mathbf{s}_i- W_j \|_2^2.
\end{equation}
Also, we use block gradient descent to obtain the solution, where the gradient is:
\begin{equation}
    \nabla_{\mathbf{W}_t} L(\mathbf{F}, \mathbf{W}) = - (W_t X_t - y_t)x_t^\top - \sum_{i=1}^T \pi^*_{i,t} (\mathbf{F} s_i - W_t)
\end{equation}

\subsection{Entropic optimal transport and Sinkhorn algorithm}
\label{apdx:sinkhorn}

\begin{algorithm2e}[h]
\caption{Sinkhorn algorithm}
\label{alg:sinkhorn}
\KwIn{Cost matrix $\mathbf{C} \in \mathbb{R}^{n \times m}$, entropy coefficient $\gamma$}
\KwOut{$\mathbf{\Pi}$}
$\mathbf{K} \xleftarrow{} \exp (- \mathbf{C} / \gamma)$, $ \mathbf{\nu} \xleftarrow{} \frac{\mathbf{1}_m}{m}$\;
\While{not converge}{
$\mathbf{\mu} \xleftarrow{}  \frac{\mathbf{1}_n}{n} \oslash \mathbf{K} \mathbf{\nu} $\;
$\mathbf{\nu} \xleftarrow{}  \frac{\mathbf{1}_m}{m} \oslash \mathbf{K}^T \mathbf{\mu} $\;}
$\mathbf{\Pi} \xleftarrow{} diag(\mathbf{\mu}) \mathbf{K} diag(\mathbf{\nu})$\;
\end{algorithm2e}

\section{Additional experiments and details}

\subsection{Uneven time}
As illustrated in figure (\ref{fig:uneven_app}), we overlay the regression results on a training set of 5 tasks where Task 5 is missing data from a significant portion of the domain.  While all multi-task regression methods, including independent models, perform fairly well on tasks 1-4,  only our proposed PhysioMTL method yields the most accurate regression result. 

\begin{figure}[h!]
\floatconts
  {fig:uneven_app}
  {\vspace{-10pt}
  \caption{Here we overlay the regression results on a training set of 5 tasks where Task 5 is missing data from a significant portion of the domain. 
  }}
  {%
    \subfigure[Task 1]{\label{fig:uneven_time_0}%
      \includegraphics[width=0.195\linewidth]{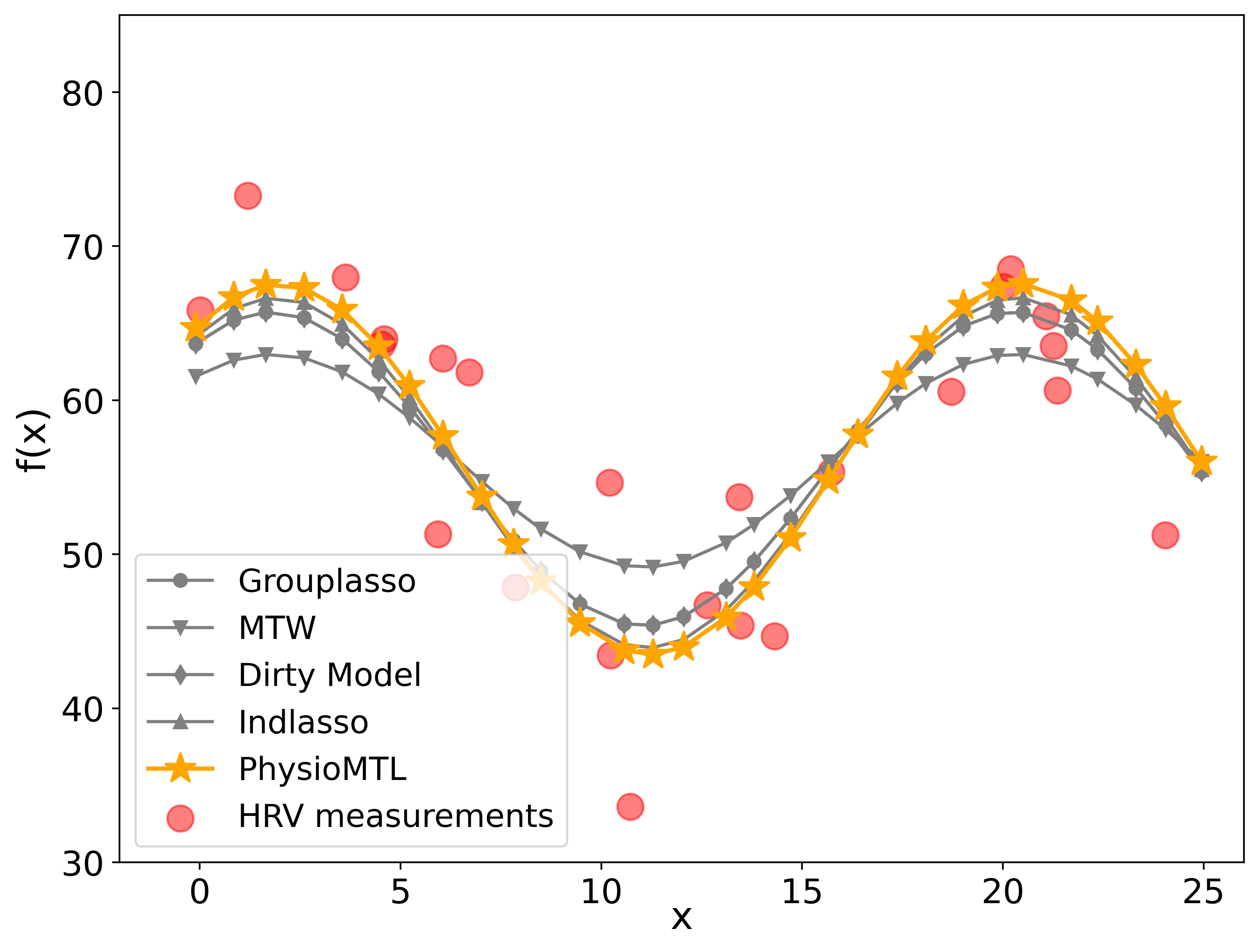}}\hfill%
    \subfigure[Task 2]{\label{fig:uneven_time_1}%
      \includegraphics[width=0.195\linewidth]{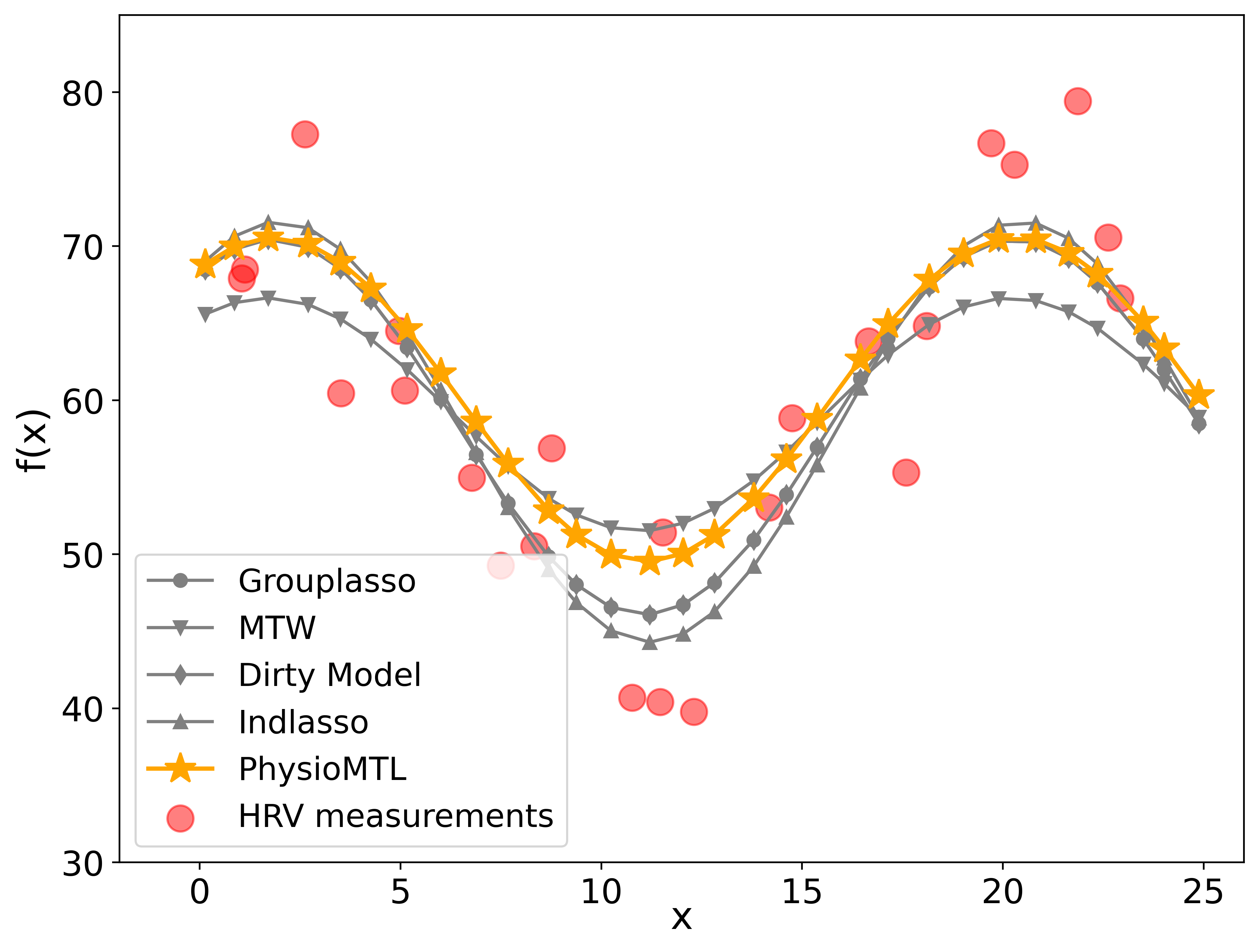}}\hfill
    \subfigure[Task 3]{\label{fig:uneven_time_2}%
      \includegraphics[width=0.195\linewidth]{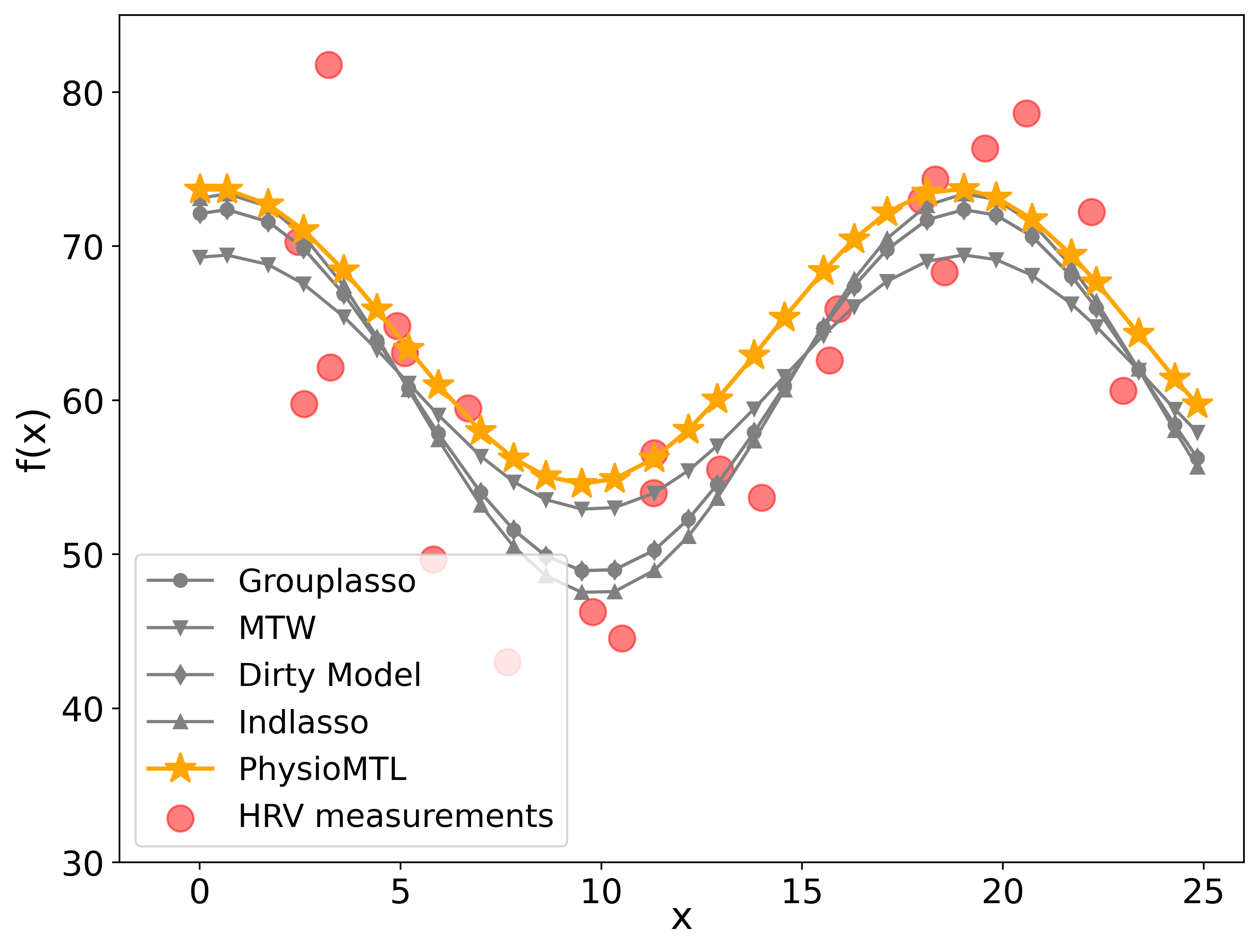}}\hfill%
    \subfigure[Task 4]{\label{fig:uneven_time_3}%
      \includegraphics[width=0.195\linewidth]{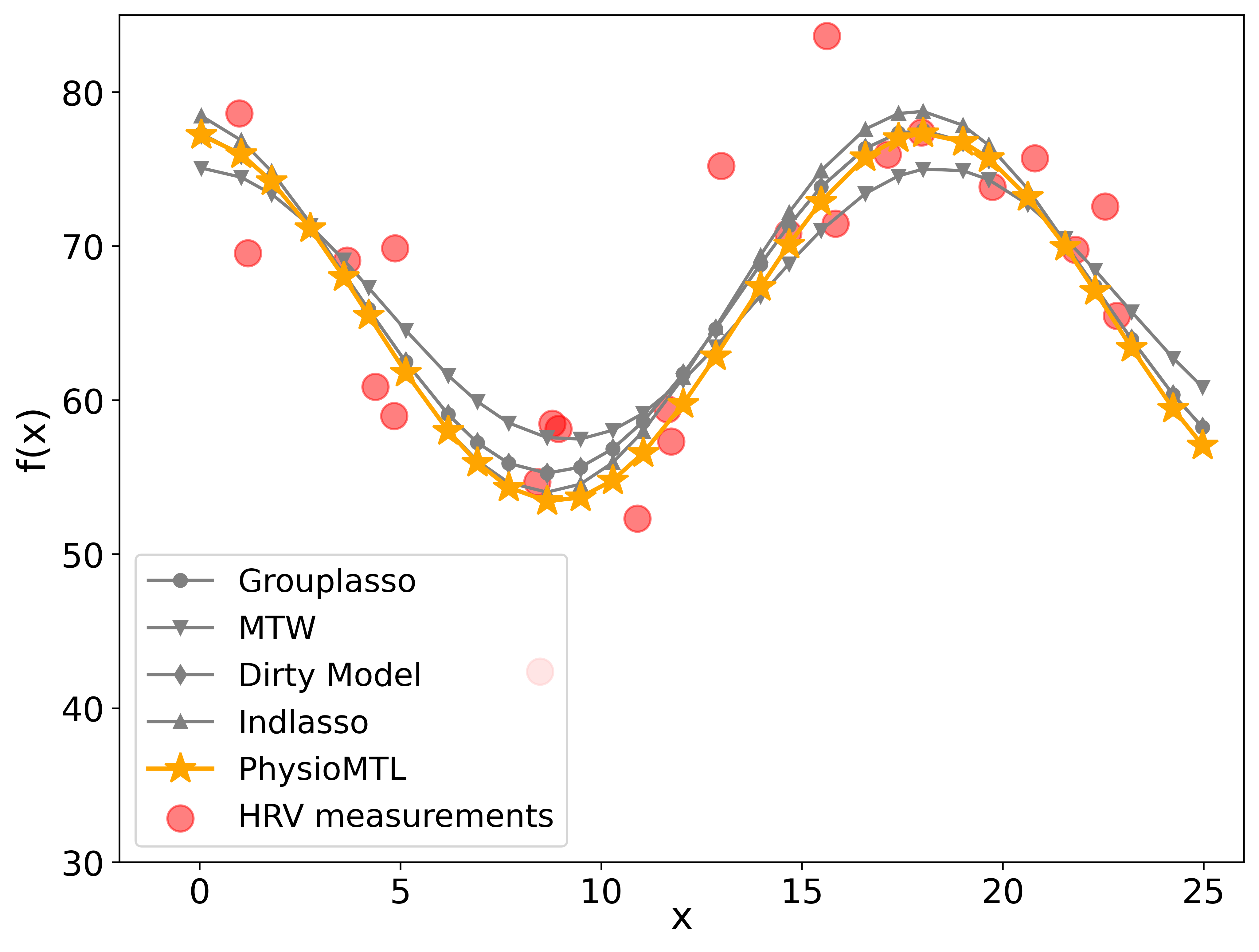}}\hfill%
    \subfigure[Task 5]{\label{fig:uneven_time_4}%
      \includegraphics[width=0.195\linewidth]{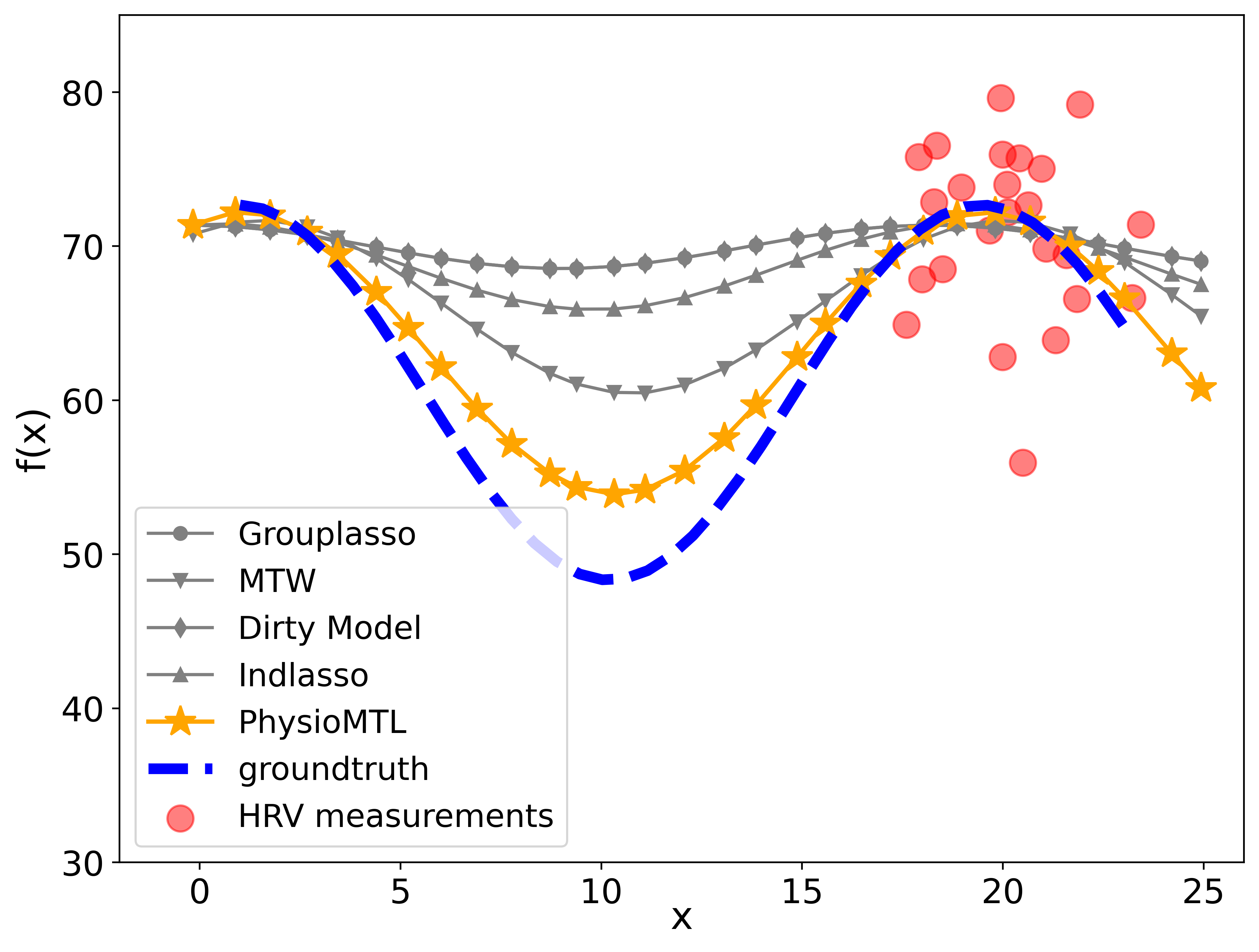}}\hfill
  } 
\end{figure}

\subsection{MMASH}

\begin{table}[h!]
\centering
\begin{tabular}{rrrrrrr}
\toprule
   & Age & BMI & Activity & Sleep & Stress  \\
\midrule
 
Mean  & 26.95  & 23.11   & 0.43 & 6.55 & 35.57  \\

Std   & 4.50   & 3.17   & 0.58 & 1.28 & 16.08 \\

Min   & 20.0   & 19.41   & 0.00 & 4.40 & 14.00 \\

25\%   & 25.0   & 21.63   & 0.00 & 5.87 & 26.00 \\

50\%   & 27.0   & 22.84   & 0.25 & 6.50 & 32.00 \\

75\%   & 28.0   & 23.18   & 0.67 & 7.27 & 41.00 \\

Max   & 40.0   & 33.24   & 2.00 & 10.50 & 74.00 \\
\bottomrule
\end{tabular}
\vspace{-5pt}
\caption{Descriptive statistics of all the demographic and lifestyle factors we use in the MMASH dataset.}
\label{table:MMASH_statistics}
\end{table}

\begin{figure}[h!]
\floatconts
  {fig:mmash_linear}
  {\vspace{-10pt}
  \caption{The counterfactual analysis on the MMASH dataset using a linear transport map. We vary the attributes of a hypothetical subjective's task-wise features and investigate the resulting HRV variational. Each feature investigated (age, BMI, activity, sleep, and stress) is continuous, however, for ease of visualization we vary each feature on a discrete grid.
  }}
  {%
    \centering
    \subfigure[Age]{\label{fig:mmash_hrv_age_linear_2}%
      \includegraphics[width=0.29\linewidth]{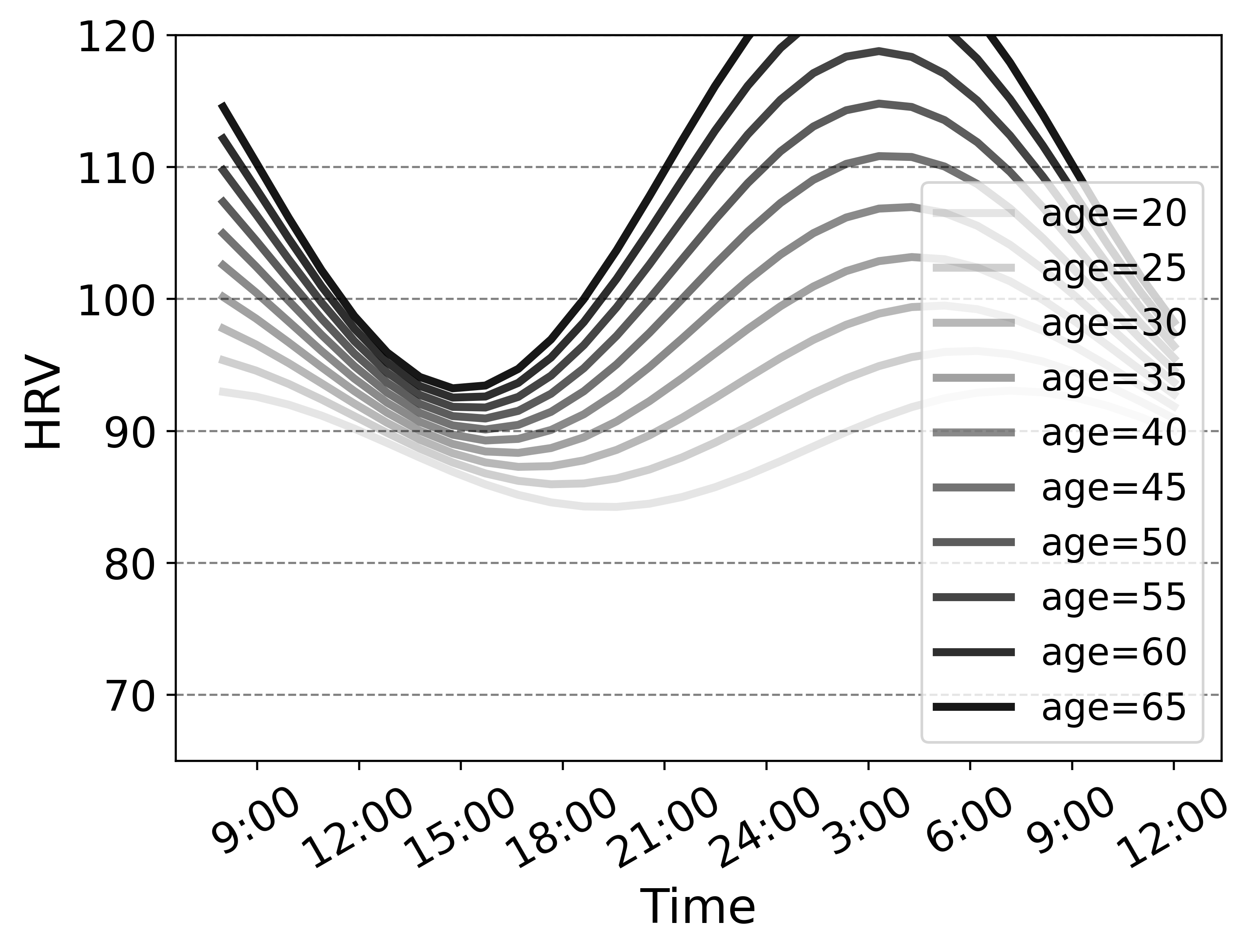}}\hfill%
    \subfigure[BMI]{\label{fig:mmash_hrv_bmi_linear_2}%
      \includegraphics[width=0.29\linewidth]{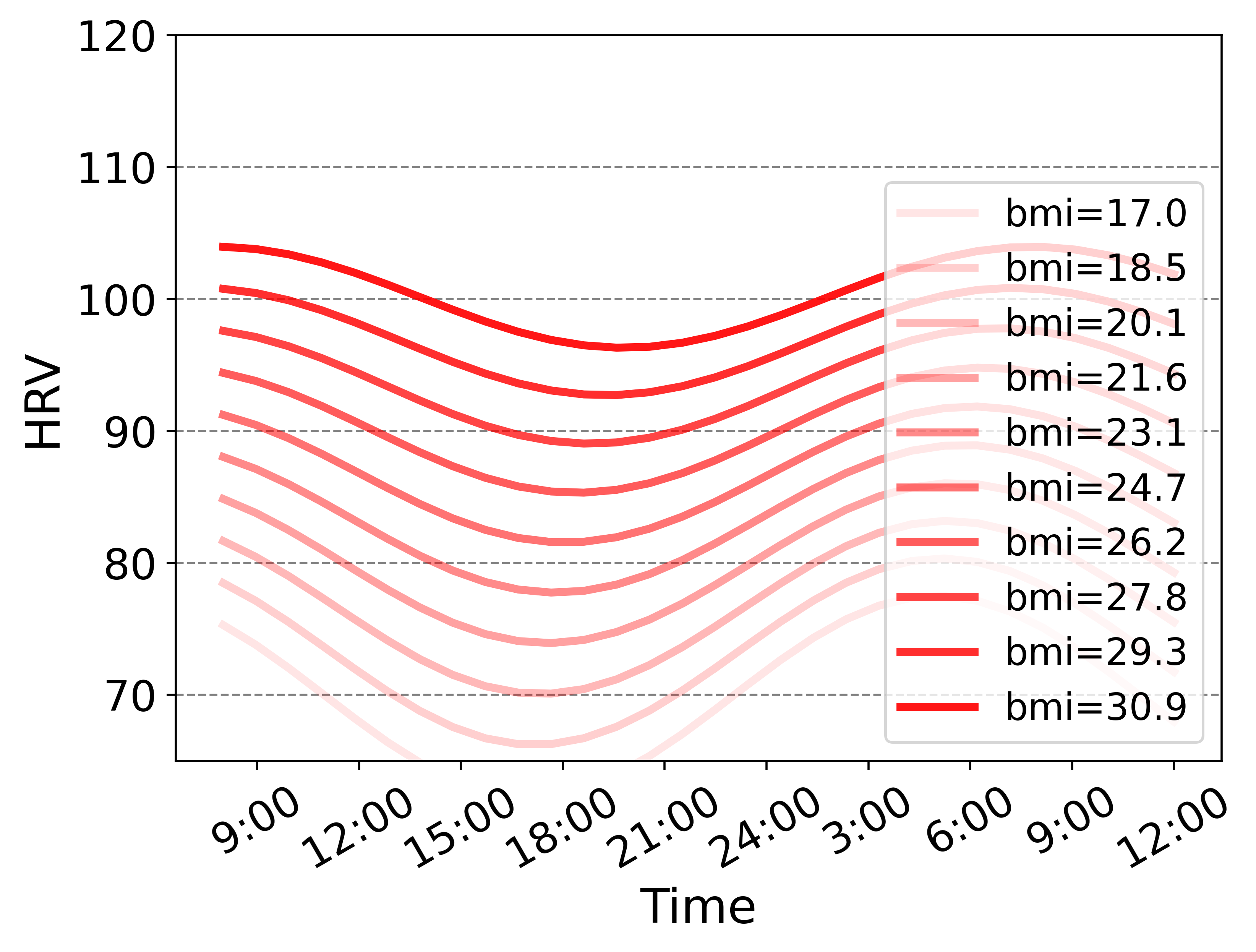}}\hfill
    \subfigure[Activity]{\label{mmash_hrv_activity_linear_2}%
      \includegraphics[width=0.29\linewidth]{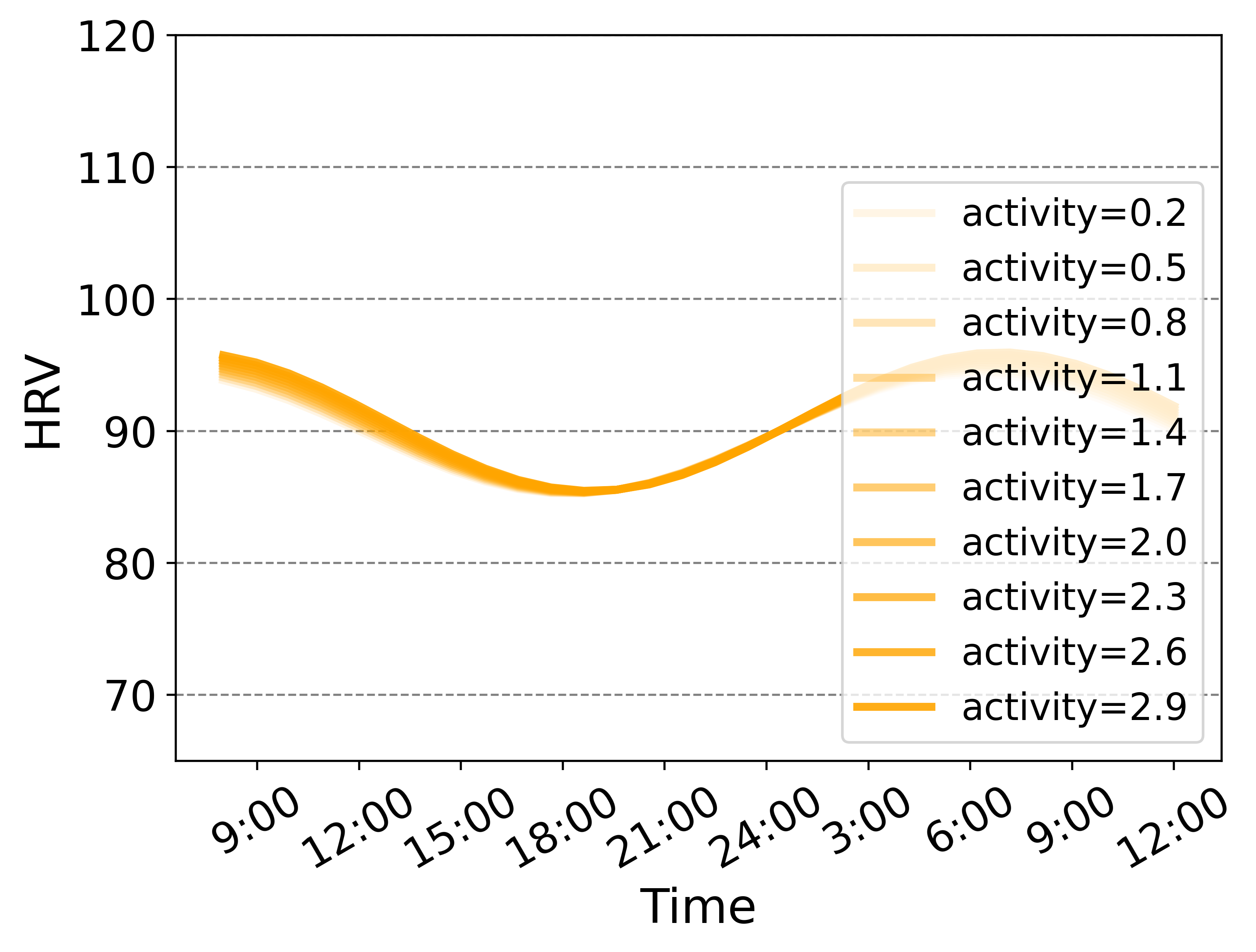}}\hfill%
    \newline
    \centering
    \subfigure[Sleep]{\label{mmash_hrv_sleep_linear_2}%
      \includegraphics[width=0.29\linewidth]{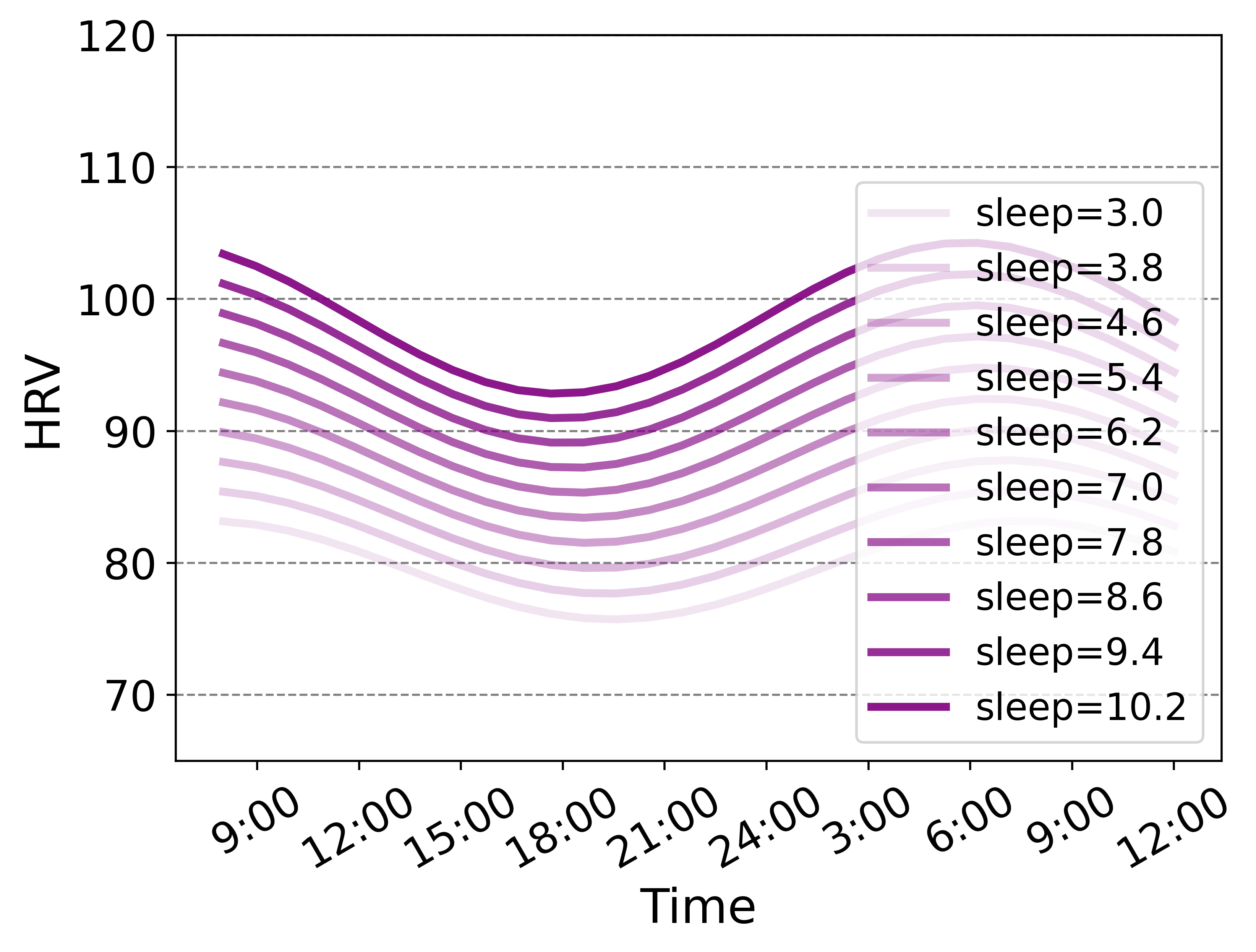}}\hfill%
    \subfigure[Stress]{\label{mmash_hrv_stress_linear_2}%
      \includegraphics[width=0.29\linewidth]{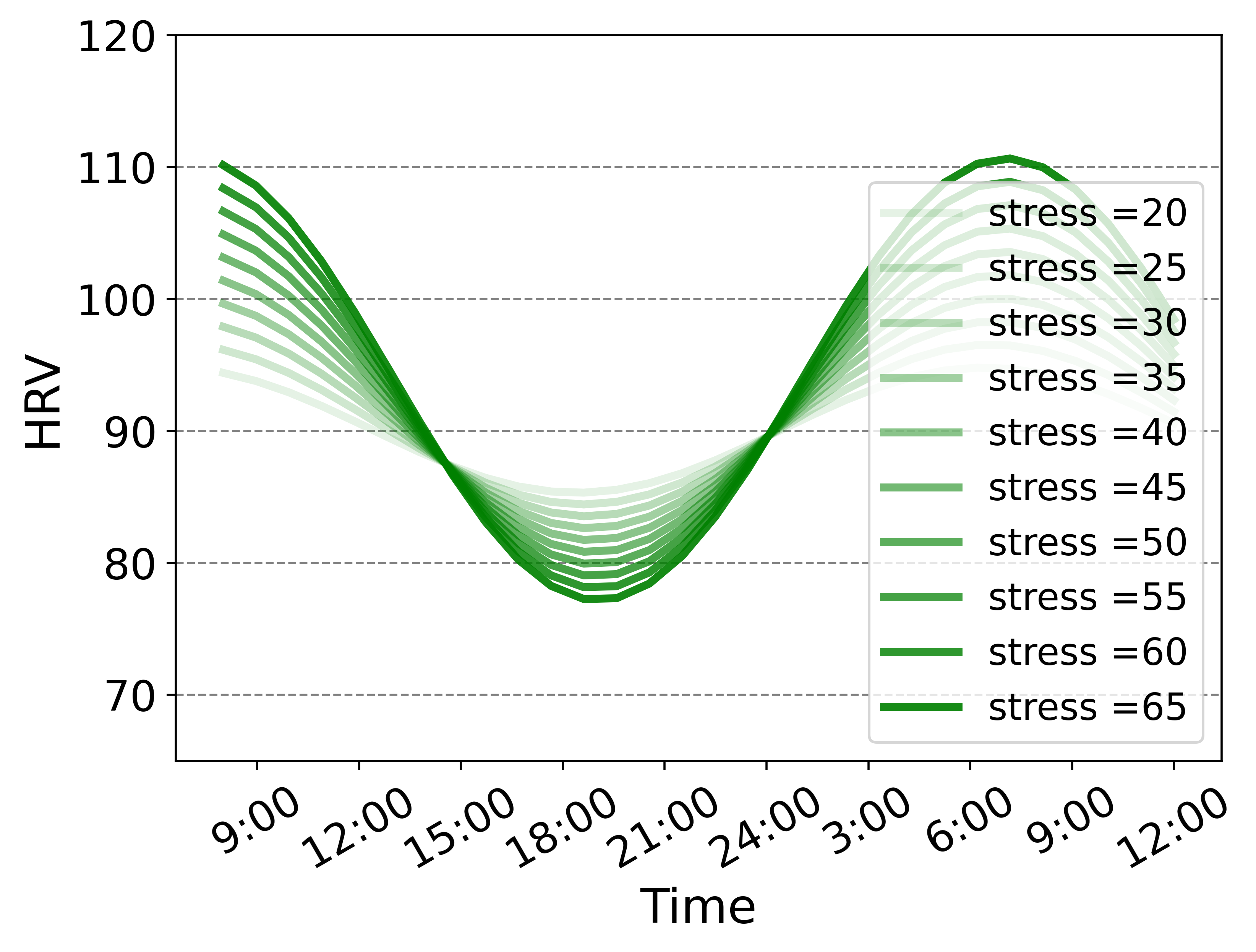}}\hfill
    \subfigure[Baseline]{\label{fig:ahms_hrv_baseline_linear_2}%
      \includegraphics[width=0.29\linewidth]{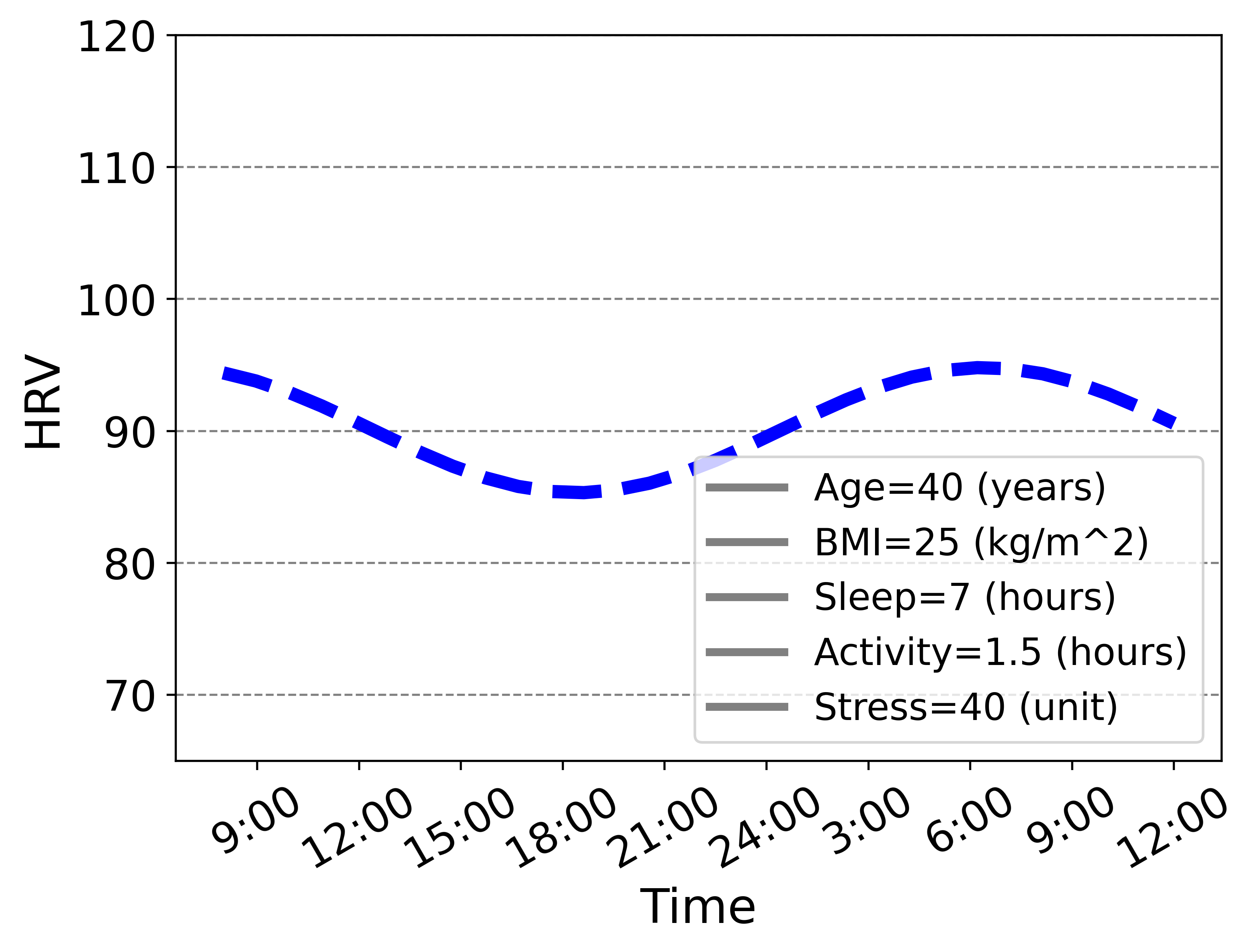}}\hfill%
  } 
\end{figure}

\subsection{AHMS}


\begin{table}[h!]
\centering
\begin{tabular}{rrrrrrr}
\toprule
   & Age & BMI & Activity & Sleep & VO2\textsubscript{max}  \\
\midrule
 
Mean  & 44.44 & 27.12  & 0.69 & 7.38 & 33.79  \\

Std   & 12.32 & 5.86  & 0.43 & 0.87 & 8.48 \\

Min   & 23.0 & 18.42 & 0.20 & 4.69 & 15.13 \\

25\%   & 35.0 & 23.09 & 0.45 & 6.95 & 26.90 \\

50\%   & 42.50 & 25.43 & 0.56 & 7.30 & 33.57 \\

75\%   & 52.25 & 30.00   & 0.79 & 7.97 & 39.28 \\

Max   & 79.0 & 47.24 & 3.21 & 10.19 & 52.60 \\
\bottomrule
\end{tabular}
\vspace{-5pt}
\caption{Descriptive statistics of all the demographic and lifestyle factors we use in the AHMS dataset.}
\label{table:AHMS_statistics}
\end{table}

\begin{figure}[h!]
\floatconts
  {fig:ahms_linear}
  {\vspace{-10pt}
  \caption{The counterfactual analysis on the AHMS dataset using a linear transport map. We vary the attributes of a hypothetical subjective's task-wise features and investigate the resulting HRV variational. Each feature investigated (age, BMI, activity, sleep, and VO2\textsubscript{max}) is continuous, however, for ease of visualization we vary each feature on a discrete grid.
  }}
  {%
    \centering
    \subfigure[Age]{\label{fig:ahms_hrv_age_linear_2}%
      \includegraphics[width=0.29\linewidth]{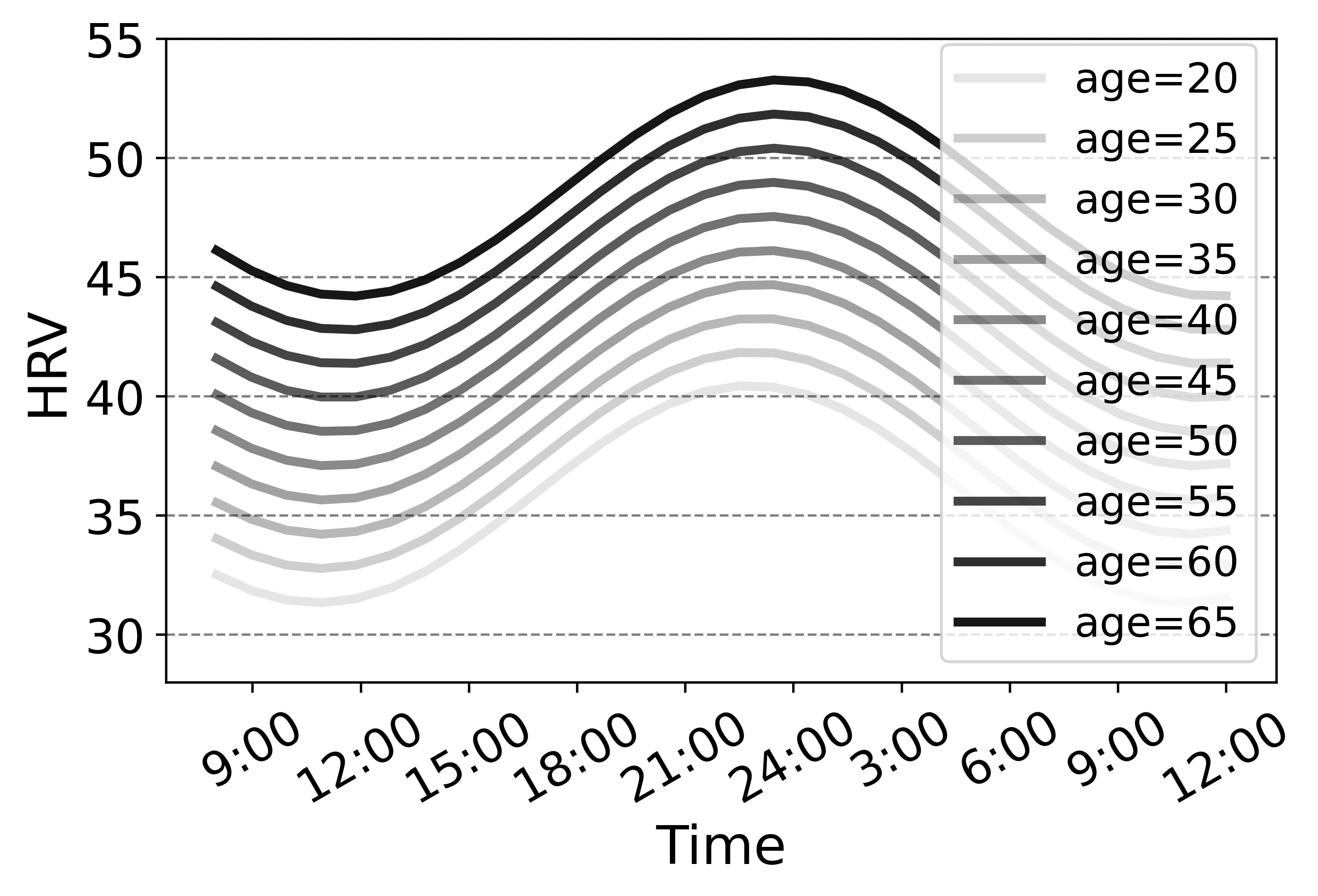}}\hfill%
    \subfigure[BMI]{\label{fig:ahms_hrv_bmi_linear_2}%
      \includegraphics[width=0.29\linewidth]{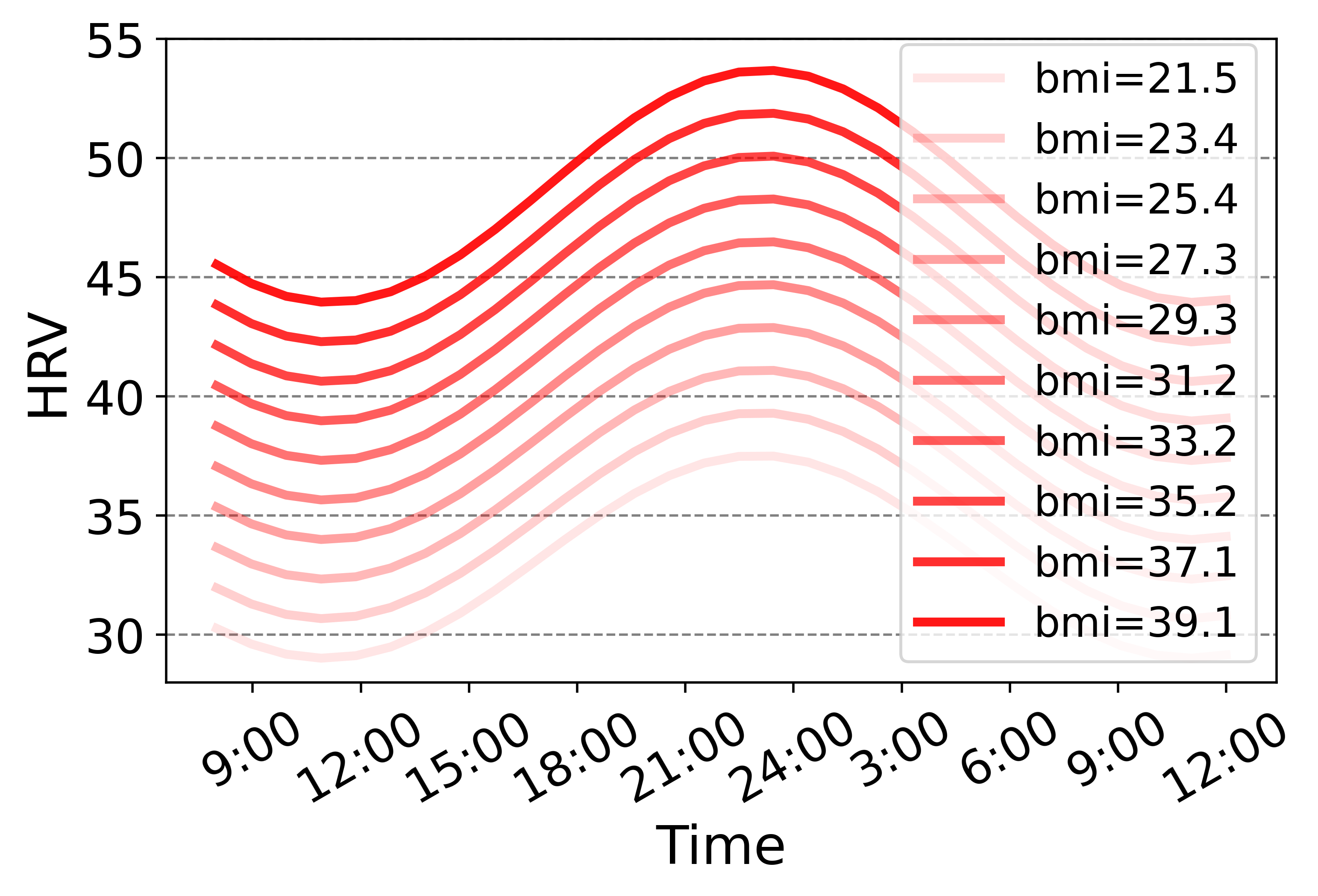}}\hfill
    \subfigure[Activity]{\label{fig:ahms_hrv_activity_linear_2}%
      \includegraphics[width=0.29\linewidth]{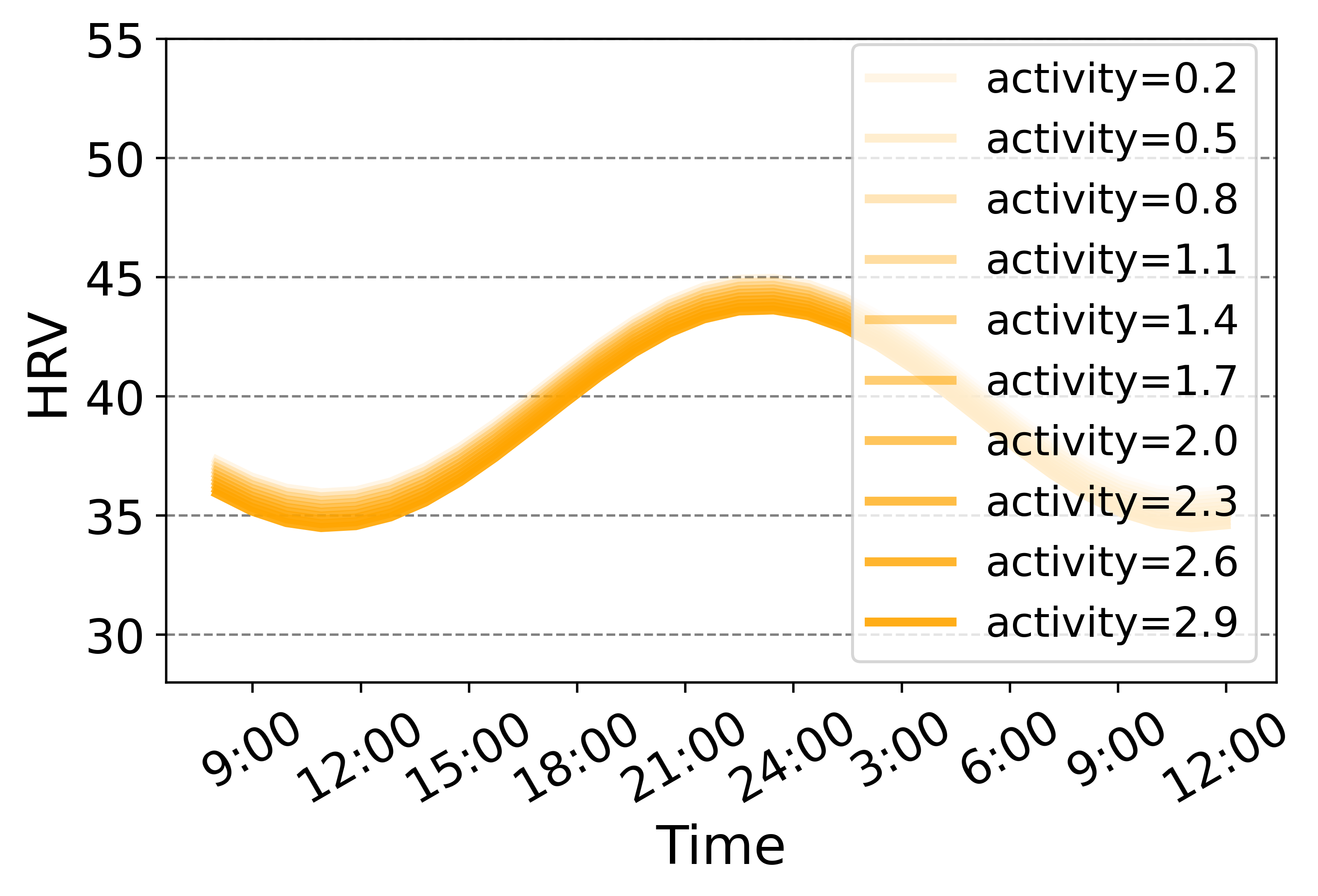}}\hfill%
    \newline
    \centering
    \subfigure[Sleep]{\label{fig:ahms_hrv_sleep_linear_2}%
      \includegraphics[width=0.29\linewidth]{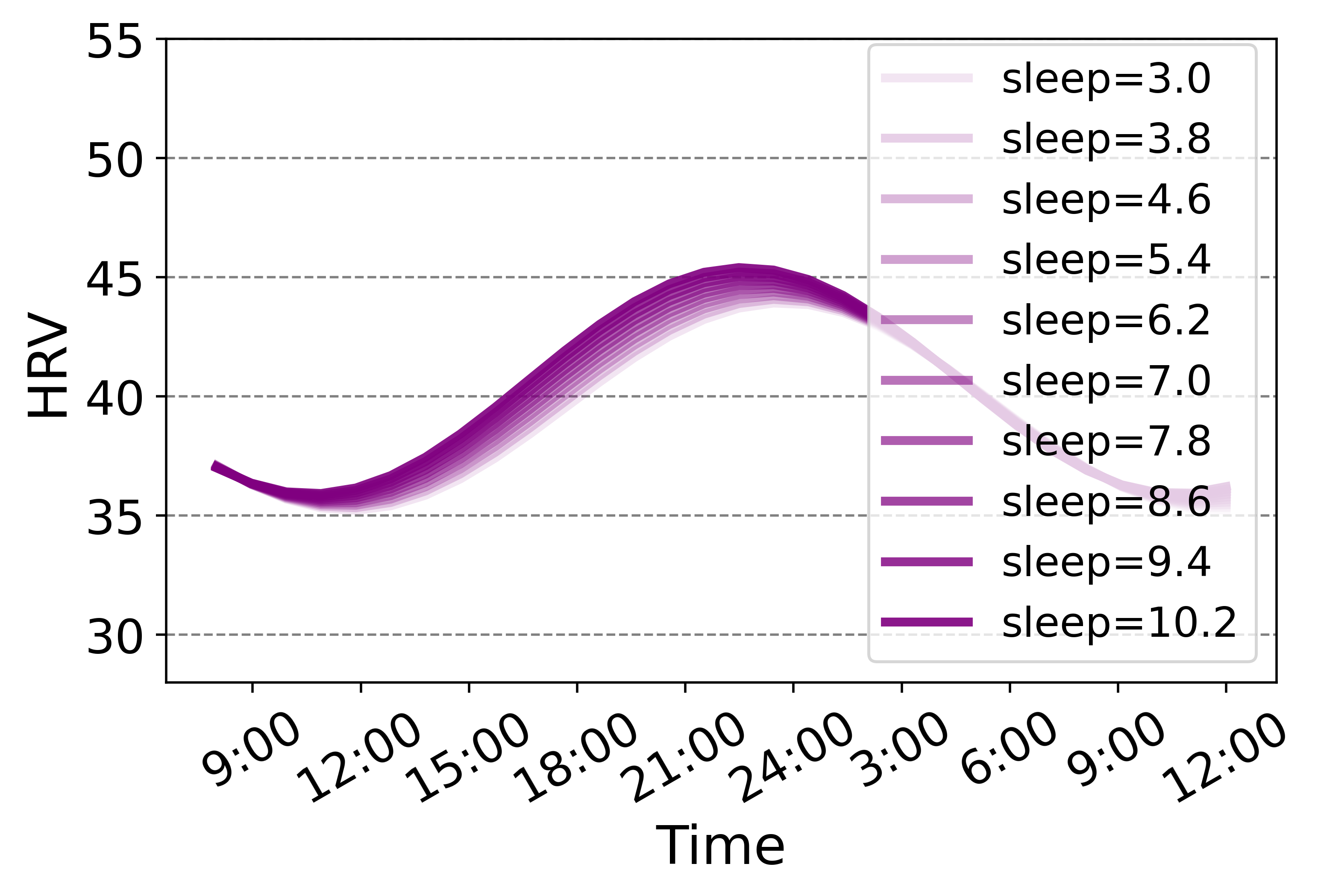}}\hfill%
    \subfigure[VO2\textsubscript{max}]{\label{fig:ahms_hrv_stress_linear_2}%
      \includegraphics[width=0.29\linewidth]{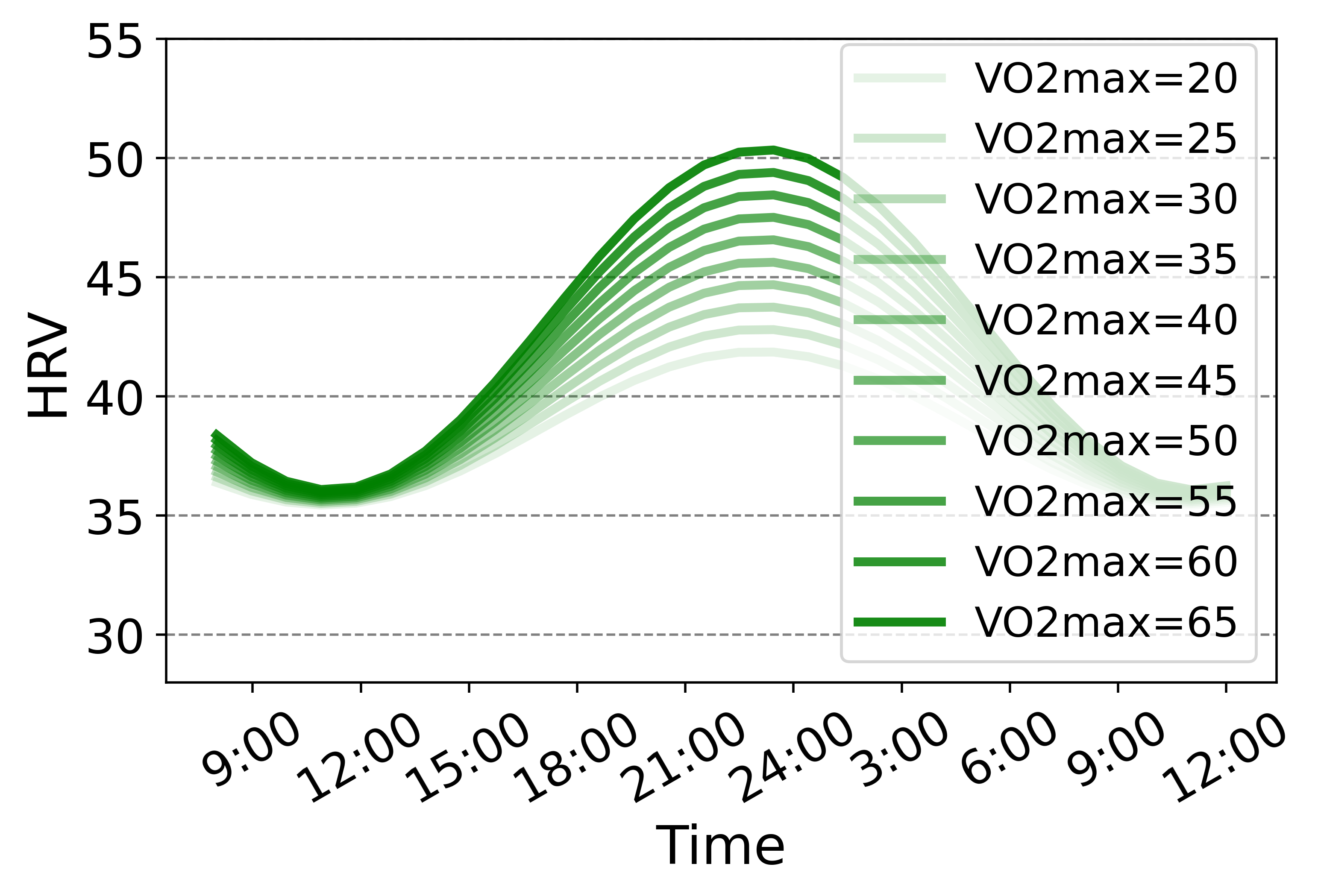}}\hfill
    \subfigure[Baseline]{\label{fig:ahms_hrv_baseline_linear_2}%
      \includegraphics[width=0.29\linewidth]{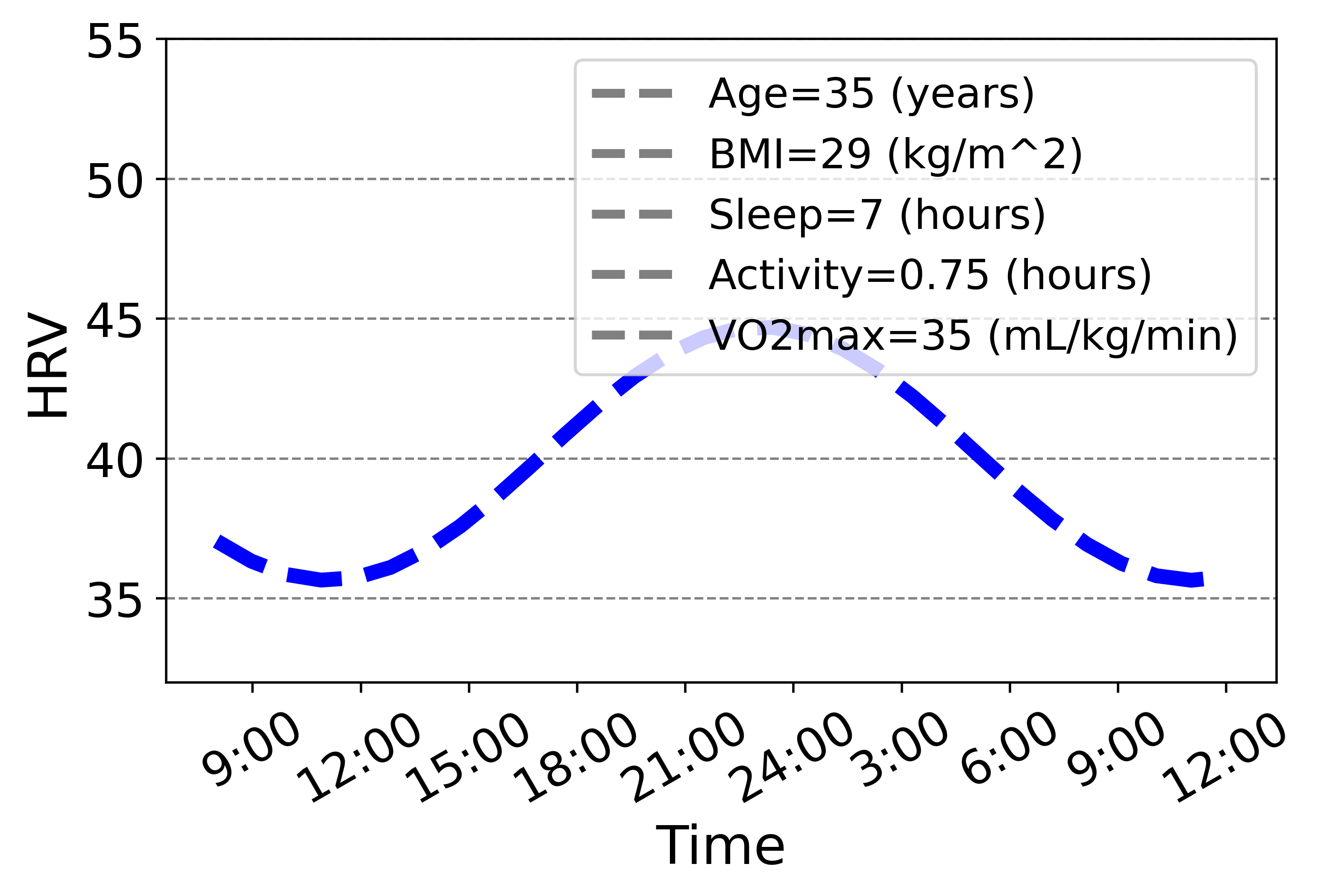}}\hfill%
  } 
\end{figure}

\end{document}